\documentclass[9pt,twocolumn]{article}
\usepackage[english]{babel}
\usepackage{rotating}
\usepackage{amsmath}
\usepackage{amsfonts}
\usepackage{geometry}
\usepackage{graphicx, subfigure}
\usepackage{verbatim}
\usepackage{color}
\usepackage{aas_macros}
\usepackage{url}
\usepackage{multirow}
\usepackage[numbers,sort&compress]{natbib}

\newcommand\ion[2]{#1{\scshape{#2}}}%

\begin{document}

\title{A High Resolution Atlas of Composite SDSS Galaxy Spectra}

\author{L\'aszl\'o Dobos$^{1}$\thanks{E-mail: dobos@complex.elte.hu}, Istv\'an Csabai$^{1,2}$, Ching-Wa Yip$^{2}$, Tam\'as Budav\'ari$^{2}$, \and Vivienne Wild$^{3} $ and  Alexander S. Szalay$^{2}$}

\maketitle

\footnotetext[1]{E\"otv\"os Lor\'and University, Department of Physics of Complex Systems, 1117 Budapest, Hungary}

\footnotetext[2]{The Johns Hopkins University, Department of Physics and Astronomy, Baltimore, MD 21218, USA}

\footnotetext[3]{Institute for Astronomy, University of Edinburgh, Royal Observatory, Blackford Hill, Edinburgh, EH9 3HJ, U.K. (SUPA)}

\label{firstpage}

\begin{abstract}
In this work we present an atlas of composite spectra of galaxies based on the data of the Sloan Digital Sky Survey Data Release 7 (SDSS DR7). Galaxies are classified by colour, nuclear activity and star-formation activity to calculate average spectra of high signal-to-noise ratio and resolution ($ S/N \simeq 132 - 4760 $ at {$ \Delta \lambda = 1 $ \AA}), using an algorithm that is robust against outliers. Besides composite spectra, we also compute the first five principal components of the distributions in each galaxy class to characterize the nature of variations of individual spectra around the averages. The continua of the composite spectra are fitted with BC03 stellar population synthesis models to extend the wavelength coverage beyond the coverage of the SDSS spectrographs. Common derived parameters of the composites are also calculated: integrated colours in the most popular filter systems, line strength measurements, and continuum absorption indices (including Lick indices). \textcolor{black}{These derived parameters are compared with the distributions of parameters of individual galaxies and it is shown on many examples that the composites of the atlas cover much of the parameter space spanned by SDSS galaxies.} By co-adding thousands of spectra, a total integration time of several months can be reached, which results in extremely low noise composites. The variations in redshift not only allow for extending the spectral coverage bluewards to the original wavelength limit of the SDSS spectrographs, but also make higher spectral resolution achievable. The composite spectrum atlas is available online at \url{http://www.vo.elte.hu/compositeatlas}.
\end{abstract}

\paragraph{keywords}
methods: statistical -- techniques: spectroscopic -- surveys -- atlases -- galaxies: statistics

\section{Introduction}

With the advance of large scale spectroscopic surveys, spectrum stacking has become a widely-used technique in extragalactic astrophysics to derive composite spectra of high signal-to-noise ratios in order to enable the analysis of spectral features that are washed away by noise in single object observations. Recent work have concentrated on averaging and analysing single classes of celestial objects, for example luminous red galaxies \citep{Eisenstein2003}, high-redshift galaxies \citep{Abraham2004}, galaxies in clusters \citep{Dressler2004}, star-forming galaxies with varying inclination \citep{Yip2010}, or quasars \citep{VandenBerk2001}.

As the spectroscopic survey of the SDSS Legacy Surveys has finished, it is time to summarize the collected information. In this work, we focus on galaxies of various types, and present a comprehensive atlas of robustly-averaged composite spectra and spectral principal components of galaxies classified by many criteria. We use the original observed spectra as they were reduced and published in SDSS DR7 \citep{dr7}.

\subsection{Overview of the classification}

Galaxies are automatically classified following a revised classification scheme based on the work of \citet{Lee2008}. In their work, galaxies were classified by colour, nuclear activity, star-formation activity and morphology.

We separate galaxies into three colour classes: red, green and blue, green refering to green valley galaxies as defined by \citet{Strateva2001}. Furthermore, galaxies are classified by star formation activity and nuclear activity into six classes: passive, measured H$\alpha$ emission, star-forming, LINER and Seyfert galaxies; the sixth category is active galaxies with star-formation signatures hereafter denoted as AGN + \ion{H}{ii}. Star-forming galaxies, LINERs and Seyferts are classified using the Baldwin-Phillips-Terlevich (BPT) diagnostic diagram \citep{Baldwin1981}, and the segregation curves determined by \citet{Kewley2001}, \citet{Kauffmann2003a} and \citet{Kewley2006}. We omit morphological classification of the galaxies for reasons detailed in Sec.~\ref{sec:class}, where we also describe the classification scheme in detail.

\subsection{Robust average and principal components}

To compute the average spectra, we use the robust statistical method of \citet{Budavari2009}. This method not only gives the robust average and variance of a large, streaming\footnote{We apply the term ``streaming'' to data that is too large to fit into the memory so iterative loading from the background storage is necessary.} set of data vectors (the spectra, in our case), but gives the first few principal components as well. Although we are primarily interested in the average spectra, the eigenspectra can help characterise the variations of objects within the subsets of the sample.

Principal Component Analysis (PCA or Karhunen--Lo\`eve Transformation) has been widely used in astronomical spectrum analysis in the last two decades. The basic idea of PCA is to derive an orthogonal basis from multidimensional data where the basis vectors point to the directions of maximum variance of the original data vectors. It was demonstrated that PCA can be successfully applied to high-dimensional optical and UV galaxy spectra for classification purposes \citep{Connolly1995}. Real-life scenarios, when dealing with noisy and gappy data where certain spectral bins are masked out due to bad observations or other reasons, require more advanced techniques \citep{Connolly1999}. An application of this robust technique to SDSS spectra was performed for galaxies and quasars by \citet{Yip2004a, Yip2004b}. A novel robust algorithm was developed by \citet{Budavari2009} which combines the robustness and gap-handling capability of the former methods with the ability to handle extraordinarily large datasets by performing the calculations iteratively. This new method significantly reduces the computational and memory requirements of the calculations and enables handling higher dimensional data (thus processing higher resolution spectra).

\subsection{Use cases of the spectrum atlas}

We would like to emphasize two use cases of this spectrum atlas. One of the keys to the success of template-based photometric redshift estimators is the good set of spectral templates used to fit the broadband spectral energy distributions of observed galaxies. Spectral templates from models are often used, but they can miss some important features of real spectra. Also, models have to be highly consistent with the observations in terms of integrated colours. As an alternative to models, individual empirical spectra are used, but it is not easy to get a high S/N set of spectra that represents all types of galaxies. Our composites can provide a high S/N, reliable and representative template set for photo-z applications.

The other important use-case of our library is generating mock catalogues. We expand the individual galaxy spectra on the eigenbases of of the various galaxy classes to determine the so called eigencoefficients. The best fitting Gaussian curves to the distribution of these eigencoefficients are determined and the parameters of the Gaussian fits ($ m $, $ \sigma $) are published, along with the eigenspectra. By using these parameters of the distributions, linear combinations of the average composite spectra and eigenspectra can be easily constructed to further extend the parameter space coverage of the atlas.

\subsection{Structure of the paper}

In Sec.~\ref{sec:data} we present the data we used for the analysis. In Sec.~\ref{sec:class} we describe our classification scheme in detail. We present the details of the averaging method and the composite spectra in Sec.~\ref{sec:composite}, where we also discuss the performance and robustness of the averaging algorithm. \textcolor{black}{In Sec.~\ref{sec:map} we fit stellar population synthesis models to the composites and present colours, emission line measurements and Lick indices of them. By comparing the derived physical parameters with the parameter distributions of individual galaxies, we show that the composites of the atlas are representative to the whole SDSS galaxy ensemble.} We summarize our work in Sec.~\ref{sec:summary}.

\section{The data}
\label{sec:data}

We analyse galaxy spectra from SDSS DR7 \citep{dr7} as reduced by the \texttt{spectro} pipeline \citep{Stoughton2002}, version 2.6. All spectra (including duplicate observations) are selected for analysis which were identified as galaxies by the \texttt{spectro} pipeline and have more than 3800 well-measured spectral bins.

Photometric data are taken from the SDSS DR7 ``best'' dataset as measured by the SDSS \texttt{photo} image reduction pipeline \citep{Stoughton2002}.

K-corrections and absolute magnitudes are taken from the New York University Value-Added Galaxy Catalogue \citep{Blanton2005}. K-corrections are calculated for $ z = 0.1 $, absolute magnitudes are based on the concordance $ \Lambda $CDM cosmological model with parameters of {$ H_0 = 100 $ km s$ ^{-1} $}, $ \Omega_M = 0.3 $ and $ \Omega_\Lambda = 0.7 $.

For line measurements of observed spectra, we use the MPA/JHU Value-Added Galaxy Catalogue \citep{Kauffmann2003b, Tremonti2004, Brinchmann2004}.

Where not otherwise noted, galaxy luminosities are quoted in Petrosian magnitudes. Wavelength values are expressed in vacuum throughout the paper.


\section{Classification scheme}
\label{sec:class}

We base our galaxy classification scheme on the work of \citet{Lee2008}, but define several additional classes as well. In the their scheme, galaxies were classified by morphology (early-type and late-type), colour (red and blue) and nuclear activity (passive, star-forming, LINER and Seyfert). We extend this scheme by classifying galaxies by the strength of the H$\alpha$ emission (completely passive galaxies and those showing some nebular emission but not enough to classify them using the BPT diagram). We also introduce a transitional class between star-forming galaxies and active galactic nuclei (AGNs) based on the works by \citet{Kewley2001, Kauffmann2003a, Kewley2006}. \citet{Kauffmann2003a} termed this latter class `composite'; we will refer to these galaxies as `AGN + \ion{H}{ii}' to avoid confusion with \textit{composite spectra}. The colour classification is extended to treat green-valley galaxies separately. Volume-limited cuts are imposed on all subsamples as described in Sec.~\ref{sec:volumes}. For the most numerous galaxy classes, we also define further classes based on colour cuts in Sec.~\ref{sec:classrefine}, and term these classes `refined star-forming' or `SF', and `refined red' or `RED'. The number of galaxies used to calculate the composites, the wavelength coverage and the estimated signal-to noise ratio are summarized in Tab.~\ref{tab:specpar}.

\begin{table}
	\begin{center}
		\begin{tabular}{ | l | r | r | r | r | }
\hline
class & $\lambda_\text{min}$ & $\lambda_\text{max}$ & count & $S/N$ \\
 & [\AA] & [\AA] &  &  \\
\hline \hline
red P & 3350 & 8900 & 10431 & 1884 \\
red H$\alpha$ & 3350 & 8950 & 61237 & 3996 \\
red SF & 3550 & 8950 & 481 & 225 \\
red A+\ion{H}{ii} & 3350 & 8950 & 726 & 329 \\
red L & 3250 & 8850 & 901 & 483 \\
red S & 3350 & 8950 & 577 & 347 \\
red all & 3350 & 8950 & 86690 & 4760 \\
\hline
green P & 3350 & 8900 & 2095 & 791 \\
green H$\alpha$ & 3350 & 8950 & 19246 & 1858 \\
green SF & 3550 & 8950 & 1533 & 482 \\
green A+\ion{H}{ii} & 3350 & 8900 & 1067 & 431 \\
green L & 3250 & 8800 & 327 & 233 \\
green S & 3350 & 8950 & 942 & 442 \\
green all & 3350 & 8950 & 26613 & 2148 \\
\hline
blue P & 3350 & 8650 & 312 & 242 \\
blue H$\alpha$ & 3350 & 8900 & 26550 & 1672 \\
blue SF & 3550 & 8950 & 14304 & 1088 \\
blue A+\ion{H}{ii} & 3350 & 8900 & 2725 & 541 \\
blue L & 3250 & 8800 & 215 & 132 \\
blue S & 3350 & 8900 & 1157 & 386 \\
blue all & 3350 & 8900 & 40504 & 1856 \\
\hline
all P & 3350 & 8900 & 12838 & 1670 \\
all H$\alpha$ & 3350 & 8950 & 107033 & 3911 \\
all SF & 3550 & 8950 & 16318 & 1035 \\
all A+\ion{H}{ii} & 3350 & 8950 & 4518 & 610 \\
all L & 3250 & 8850 & 1443 & 470 \\
all S & 3350 & 8950 & 2676 & 569 \\
all all & 3350 & 8950 & 153807 & 4464 \\
\hline
RED 1 & 3350 & 8950 & 17913 & 1777 \\
RED 2 & 3350 & 8950 & 22934 & 2458 \\
RED 3 & 3350 & 8950 & 46470 & 3939 \\
RED 4 & 3350 & 8950 & 17546 & 2303 \\
RED 5 & 3350 & 8950 & 2469 & 629 \\
\hline
SF 1 & 3550 & 9000 & 301 & 192 \\
SF 2 & 3550 & 9000 & 2862 & 650 \\
SF 3 & 3550 & 9050 & 7073 & 1000 \\
SF 4 & 3550 & 9000 & 4436 & 798 \\
SF 5 & 3550 & 9000 & 1948 & 513 \\
\hline
\end{tabular}

	\end{center} 
	\caption{Basic parameters of the composites by activity and colour class.}
	\label{tab:specpar}
\end{table} 

In contrast to \citet{Lee2008}, we omit morphological classification of the galaxies for the following reasons. Morphological classification relies on the radial colour gradient of galaxies which was not measured perfectly by the SDSS \texttt{photo} pipeline for the whole \texttt{spectro} sample. Thus, galaxy classes defined by morphology would contain about $60$ per cent fewer galaxies: only those with good radial colour gradient measurements. Using morphological information would reduce the number of the classified galaxies to a number that would not allow reaching the desired low noise level and resolution. Also, the size and magnitude dependent incompleteness of the morphological classification would introduce unwanted bias into the data set.

We further explain the classification criteria in detail in the following subsections.

\subsection{Classification by colour}
\label{sec:colorcut}

We organise the galaxies into three colour classes based on their redshifts and $g-r$ colours. The three colour classes are the following:

\begin{itemize}

\item Red galaxies: Having primarily old stellar populations or galaxies with significant intrinsic extinction.

\item Green galaxies: Those belonging to the green valley, thought to form the transitional region between active star-formation and passivity \citep{Strateva2001, Bell2004, Baldry2004}.

\item Blue galaxies: Mostly actively star-forming galaxies.

\end{itemize}

  \begin{figure}
    \begin{center}
      \input{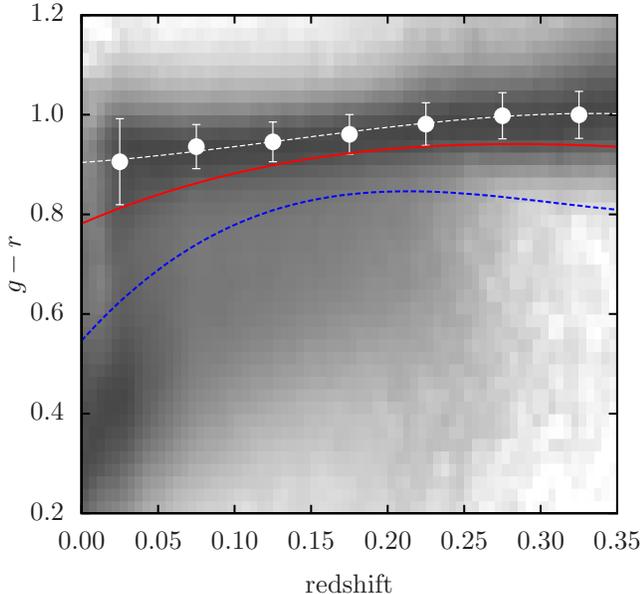}
      \vspace{24pt}
	  \caption{Definitions of the colour cuts. Density plot: logarithm of the distribution of the galaxy sample, normalised in each redshift bin. Red galaxies: above the solid red line; blue galaxies: below the dashed blue line; green valley galaxies: between the two coloured lines. The data points and the dashed white curve represent the peak of the red sequence and our best fit to them as a function of redshift. See online edition for colour version of this plot.}
       \label{fig:colorcutfig}
    \end{center}
  \end{figure} 

Fig.~\ref{fig:colorcutfig} displays the colour classes we define. Galaxies above the solid red curve are classified as `red', galaxies below the blue curve are classified as `blue' and green galaxies lie inbetween the two. These curves were determined by fitting Gaussians to the distribution of galaxies by the $g - r$ colour in bins of $\Delta z = 0.05$ between $0.0 \leq z \leq 0.35$. The dashed white line is the center of the red sequence, the solid red line is the $1.2 \sigma$ envelope of it, and the blue line is at $3.5 \sigma$ from the red sequence. To construct this plot, all galaxies were used independently of their apparent or absolute magnitudes.

In Fig.~\ref{fig:colorcutfig}, colour curves are defined by the $\sigma$ parameter of the fitted Gaussians as explained above. Since the photometric error increases with increasing redshift, it makes the scatter in the red sequence higher. This causes the colour curves to turn over around $z \simeq 0.2$ and be more restrictive for green-valley and blue galaxies, allowing more variance in the colour of red galaxies. Because of the relatively low number of observed green-valley and blue galaxies at these redshifts, this would have no effect on the average composites, but more high-redshift red galaxies could be used for the analysis. In the present analysis we set the redshift limits of the subsample volumes below $z \leq 0.17$.

\subsection{Classification by activity}

Our classification scheme by star-formation and nuclear activity is based on the works by \citet{Kewley2001} and \citet{Kauffmann2003a}. LINER/Seyfert separation follows \citet{Kewley2006}. We define the following six classes based on activity:

\begin{itemize}

\item Passive galaxies: No measurable H$\alpha$ emission.

\item H$\alpha$ measured: Hydrogen emission is measured, but other lines are not strong enough for classification.

\item Star-forming galaxies: with clear star-formation signatures but no active nucleus.

\item AGN + \ion{H}{ii} galaxies: with AGN and extreme starburst signatures at the same time.

\item LINER galaxies: pure LINER signatures \citep{Heckman1980}.

\item Seyfert galaxies: pure Seyfert signatures.

\end{itemize}

\begin{figure}
    \begin{center}
      \input{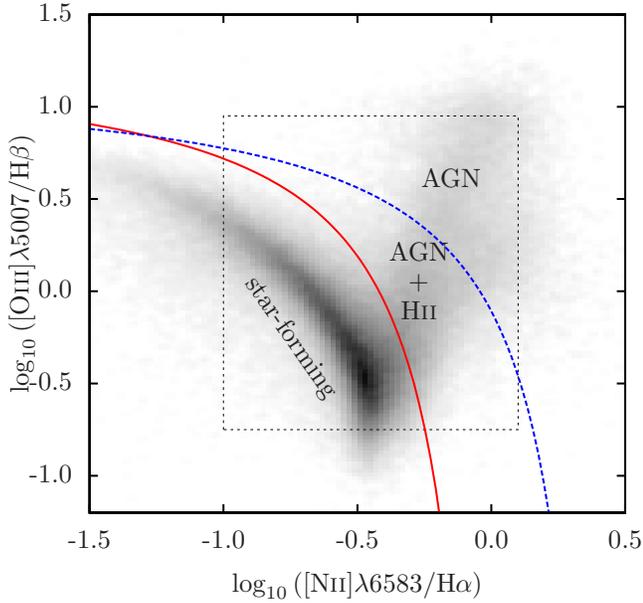}
      \vspace{10pt}
      \caption{BPT diagram for our sample of SDSS galaxies. The density plot is the square root of the number density of SDSS galaxies with high enough line S/N measurements (see text). The solid red curve indicates the segregation line between star forming galaxies and active galactic nuclei. The blue dashed curve is the extreme starburst line. The dashed rectangle marks the same area as displayed on Fig.~\ref{fig:bptcomposites}. See text for references.}
      \label{fig:bpt}
    \end{center}
  \end{figure}

In Fig.~\ref{fig:bpt} we plot the BPT diagram of the galaxies in our sample. The segregation line (solid red curve) between star-forming galaxies (below the curve) and AGN (above the curve) is:

  \begin{equation*}
	\log_{10}{ \frac{[\mbox{\ion{O}{iii}}]}{\mbox{H}\beta} } =
    0.61 \left[ \log_{10}{ \frac{[\mbox{\ion{N}{ii}}]}{\mbox{H}\alpha} } - 0.05 \right]^{-1} + 1.3.
  \end{equation*}
  
The limit of extreme starburst (blue dashed line) is determined as:

  \begin{equation*}
	\log_{10}{ \frac{[\mbox{\ion{O}{iii}}]}{\mbox{H}\beta} } <
    0.61 \left[ \log_{10}{ \frac{[\mbox{\ion{N}{ii}}]}{\mbox{H}\alpha} } - 0.47 \right]^{-1} + 1.19.
  \end{equation*}
  
The latter line separates pure AGN (above the curve) from AGN showing star-formation signatures (between the two curves).
  
\begin{figure}
    \begin{center}
      \input{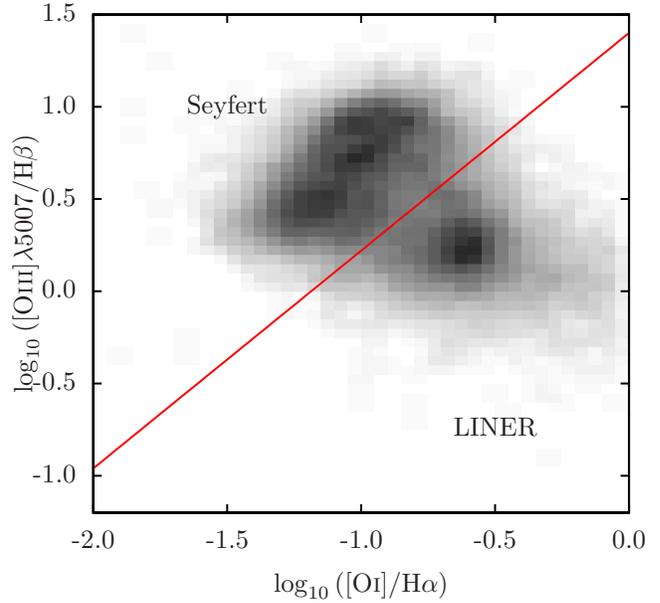}
      \vspace{10pt}
      \caption{Seyfert/LINER separation. The density plot is the square root of the number density of our sample of SDSS galaxies classified as AGN (without star-formation signature) using Fig.~\ref{fig:bpt}. The line separates the two sides of the bimodal distribution. See text for references.}
      \label{fig:bpt2}
    \end{center}
  \end{figure}

Fig.~\ref{fig:bpt2} shows the distribution of galaxies classified as AGN (without star-formation signatures). We apply the following cut to distinguish LINERs (below the line) from Seyfert galaxies (above the line):

  \begin{equation*}
	\log_{10}{ \frac{[\mbox{\ion{O}{iii}}]}{\mbox{H}\beta} } =
	1.18 \; \log_{10}{ \frac{[\mbox{\ion{O}{i}}]}{\mbox{H}\alpha} } + 1.4
  \end{equation*}
  
Only galaxies with H$\alpha$ measured at $S/N \geq 3$ are classified as star-forming, while only those with all H$\alpha$, H$\beta$, [\ion{O}{i}], [\ion{O}{iii}] and [\ion{N}{ii}] measured at $S/N \geq 3$ are classified as AGN (either AGN + \ion{H}{ii}, LINER or Seyfert). For those galaxies that do not have strong enough emission lines to make it onto the BPT, we define two further classes: completely passive with no detectable emission lines and passive with measurable H${\alpha}$.

Although their number is negligible, we excluded galaxies with extreme values of line ratios, i.e those that are outside the limits of the plot in Fig.~\ref{fig:bpt}.

\subsection{Volume-limited samples}
\label{sec:volumes}

In order to avoid the Malmquist bias, we define redshift and absolute magnitude ranges for every galaxy class, trying to maximize the number counts of the volume limited samples.

Although setting restrictive constraints on the redshift and absolute magnitude reduces the wavelength coverage of the composites, it is necessary to get physically meaningful results. The spectroscopic targeting algorithms of SDSS \citep{Eisenstein2001, Strauss2002, Stoughton2002} set the sharp faint-end $ r $-band magnitude limit at $17.7$ mag for the main galaxy sample. On the other hand, bright-end incompleteness is much less well controlled. One of the main reasons for this incompleteness is that nearby galaxies are too big in apparent diameter compared to the spectroscopic fibre diameter. We estimate the bright-end apparent $ r $-band magnitude limit to be around of $ 13.5 $ when determining the volumes.

\begin{figure*}
   \begin{minipage}{1\textwidth}
    \begin{center}
	  \vspace{24pt}
	  \hspace{18pt}
      \input{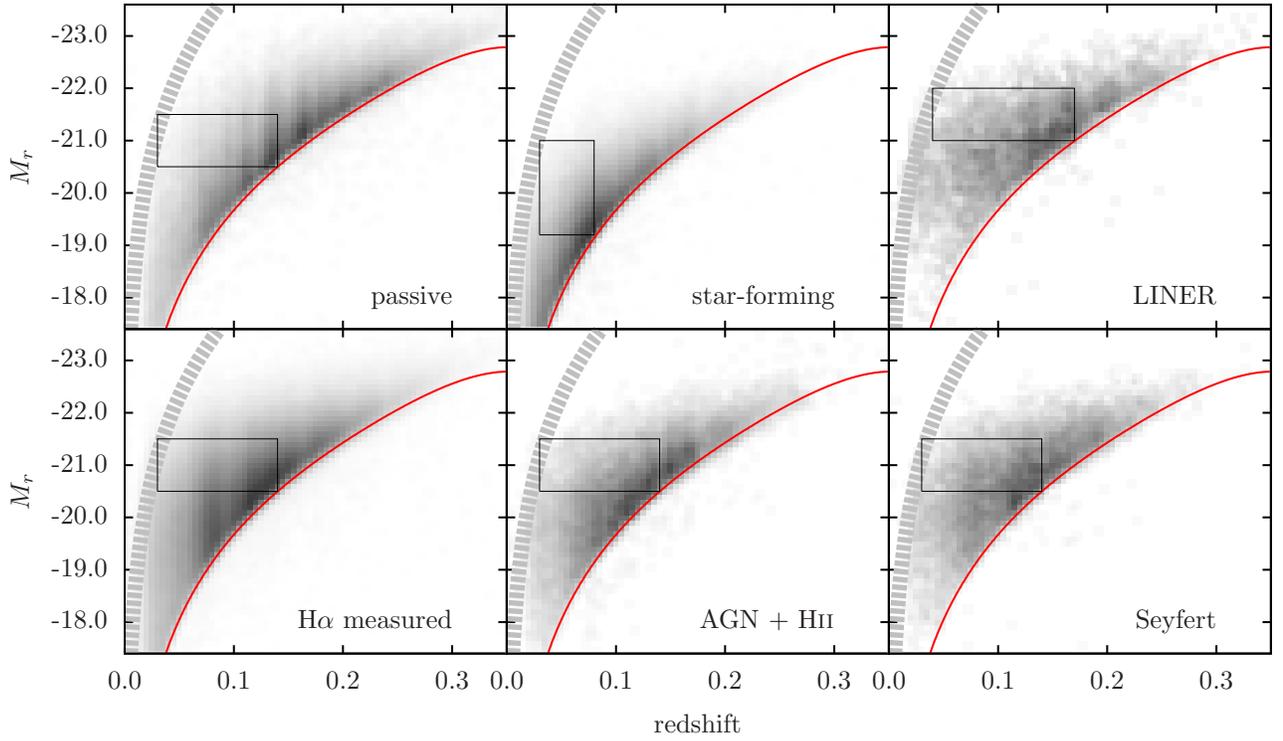}
      \vspace{36pt}
\caption{Definitions of redshift and $ r $-band absolute magnitude $ M_r $ cuts for every galaxy activity class. Density plot: square root of the galaxy density within the activity class; solid red line: faint-end magnitude limit corresponding to 17.77 mag, gray line: estimated bright-end magnitude limit corresponding to about 13.5 mag. The cuts were manually set to contain as many galaxies as possible considering the magnitude limits of the SDSS \texttt{spectro} sample.}
      \label{fig:volumes}
    \end{center}
   \end{minipage}
  \end{figure*}

We set the volumes as plotted in Fig.~\ref{fig:volumes} and summarized in Tab.~\ref{tab:volumes}.

\begin{table}
	\begin{center}
	\begin{tabular}{ | l | c | c | c | c |}
		\hline
		& $z_{\textrm{min}}$ & $ z_{\textrm{max}} $ & $ M_{r,\textrm{min}} $ & $ M_{r,\textrm{max}} $ \\
		\hline
		\hline
		passive &            0.03 & 0.14 & -20.5 & -21.5 \\
		H$\alpha$ measured & 0.03 & 0.14 & -20.5 & -21.5 \\
		star-forming &       0.03 & 0.08 & -19.2 & -21.0 \\
		AGN + \ion{H}{ii} &  0.03 & 0.14 & -20.5 & -21.5 \\
		LINER &              0.04 & 0.17 & -21.0 & -22.0 \\
		Seyfert &            0.03 & 0.14 & -20.5 & -21.5 \\
		\hline		
	\end{tabular}
	\caption{Redshift and absolute magnitude ranges for each galaxy activity class.}
	\label{tab:volumes}
	\end{center}
\end{table}
  
\subsubsection{Effect of incomplete sampling}

We note that the continua of average composite spectra are physical in the sense that they represent a non-negative linear combination of real stellar populations. This makes it meaningful to fit composites with theoretical population synthesis models. However, incomplete sampling of galaxies for composite calculation has the effect of making the composites non-physical.

\begin{figure}
    \begin{center}
      \input{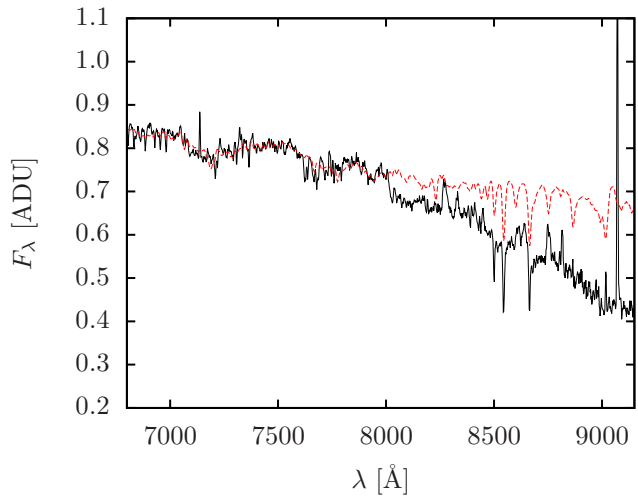}
      \vspace{10pt}
      \caption{The effect of incomplete sampling. Using low redshift spectra would allow extending the composites redwards but bright-end incompleteness causes the averaged spectrum to turn down at longer wavelength making the composite non-physical. Solid black line: composite spectrum calculated from an incomplete sample, dashed red line: best fit model. (Continuum fitting was done in the $ 3500 $ -- $ 9000 $ \AA \ range).}
      \label{fig:incomplete}
    \end{center}
  \end{figure}

While we consider spectral evolution negligible in the small redshift intervals we choose, care must be taken about the dependence of spectra on the absolute magnitude of the objects. For instance, choosing a lower redshift cut might enable extending the wavelength coverage of the composites redwards, but bright-end incompleteness would lead to the dominance of lower luminosity galaxies which (in case of blue galaxies, for example) tend to be bluer. This would result in a downturn of the composite spectrum as it can be seen in Fig.~\ref{fig:incomplete}.

\subsection{Refined classification of blue and red galaxies}
\label{sec:classrefine}

Template-based photometric redshift (photo-z) estimation algorithms (for a recent review, see \citet{PHAT2010} and references therein) rely on good quality empirical galaxy spectra or theoretical models to fit the broadband observations with; though certain techniques exist to improve the reliability of the models \citep{Budavari2000}. Most often, the problem with using theoretical models for photo-z is the mismatch between the colours calculated from the models and the colours of the observed galaxies. Also, photo-z techniques are usually based on a small number of templates in order to reduce computational complexity and to avoid spurious results due to fitting too many free parameters of the not-so-reliable models.

The most numerous galaxies (passive red ellipticals and star-forming blue spirals) show significant variations in their colours. Using only a few templates might cause the photo-z estimates to be coarse because the templates are unable to characterize this intrinsic colour variance of the objects. More reliable empirical templates, however, could be used to refine photometric redshift estimation. In order to address this issue, we further refine the most numerous galaxy classes by colour and calculate their composite spectra.

\subsubsection{Blue galaxies}

In the case of star-forming galaxies (mostly spirals and irregulars), the intrinsic colour variance is caused by many effects. Star-formation rate is primarily responsible for the restframe colours of galaxies, while dust content and inclination play an important role too \citep{Yip2010}. Also, since fibre spectrographs with limited aperture only observe the central, redder region of the galaxies, the distance of the galaxies and their bulge-to-disc ratio is an important factor causing colour variations in small aperture magnitudes. We will denote these classes as `SF' followed by a number.

\subsubsection{Red galaxies}

In the case of red galaxies (mostly ellipticals), the origin of the colour variation is mainly an evolutionary effect caused by their passively evolving old stellar populations (for a comprehensive review, see \citet{Renzini2006} and references therein); the effect of dust content on their optical colours is thought to be minimal.

For the purposes of refined colour classification, we will use all galaxies of the former `passive' and `H$\alpha$ measured' classes and denote these class as `RED' followed by a number.

\subsubsection{The refined colour classes}

\begin{figure}
    \begin{center}
      \input{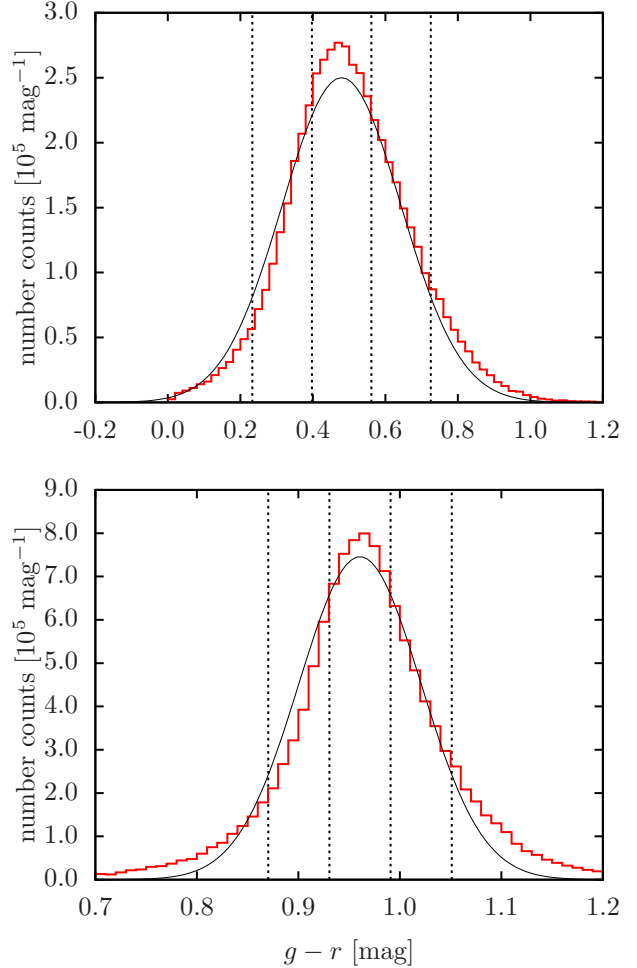}
      \caption{$g - r$ colour distribution of star-forming galaxies (top panel) and passive galaxies (bottom panel). The red histograms show the measured distributions; black curves indicate the fitted Gaussians. The vertical dashed lines mark the colour cuts we defined at $ 0.5\sigma $ and $ 1.5\sigma $ distances from the mean.}
      \label{fig:refinedcolorcut}
    \end{center}
  \end{figure}

We define the refined colour classes based on the $g - r$ colour distributions of star-forming and passive galaxies as plotted in Fig.~\ref{fig:refinedcolorcut}. The histograms of the distributions are fitted with Gaussians and the colour limits are defined at $ 0.5\sigma $ and $ 1.5\sigma $ distances from the mean. Altogether, five additional colour classes are defined, as listed in the following table:

\begin{center}
	\begin{tabular}{ | l || c | }
		\hline
		\# & limits \\
   		\hline \hline
   		1 & $ g - r < m - 1.5\sigma $ \\
   		2 & $ m - 1.5\sigma \leq g - r < m - 0.5\sigma $ \\
   		3 & $ m - 0.5\sigma \leq g - r \leq m + 0.5\sigma $ \\
   		4 & $ m + 0.5\sigma < g - r \leq m + 1.5\sigma $ \\
   		5 & $ m + 1.5\sigma < g - r  $ \\ 		
  	   	\hline  
	\end{tabular}
\end{center}

Parameters $ m $ and $ \sigma $ denote the mean and standard deviation of the fitted distributions, respectively. The fitted parameters are summarised in the following table:

\begin{center}
	\begin{tabular}{ | l || c | c | }
		\hline
		 & $ m $ & $ \sigma $ \\
   		\hline \hline
   		SF & $ 0.48 $ & $ 0.16 $ \\
   		RED & $ 0.95 $ & $ 0.06 $ \\
  	   	\hline  
	\end{tabular}
\end{center}

\subsection{Redshift distribution}

\begin{figure*}
   \begin{minipage}{1\textwidth}
    \begin{center}
	  \input{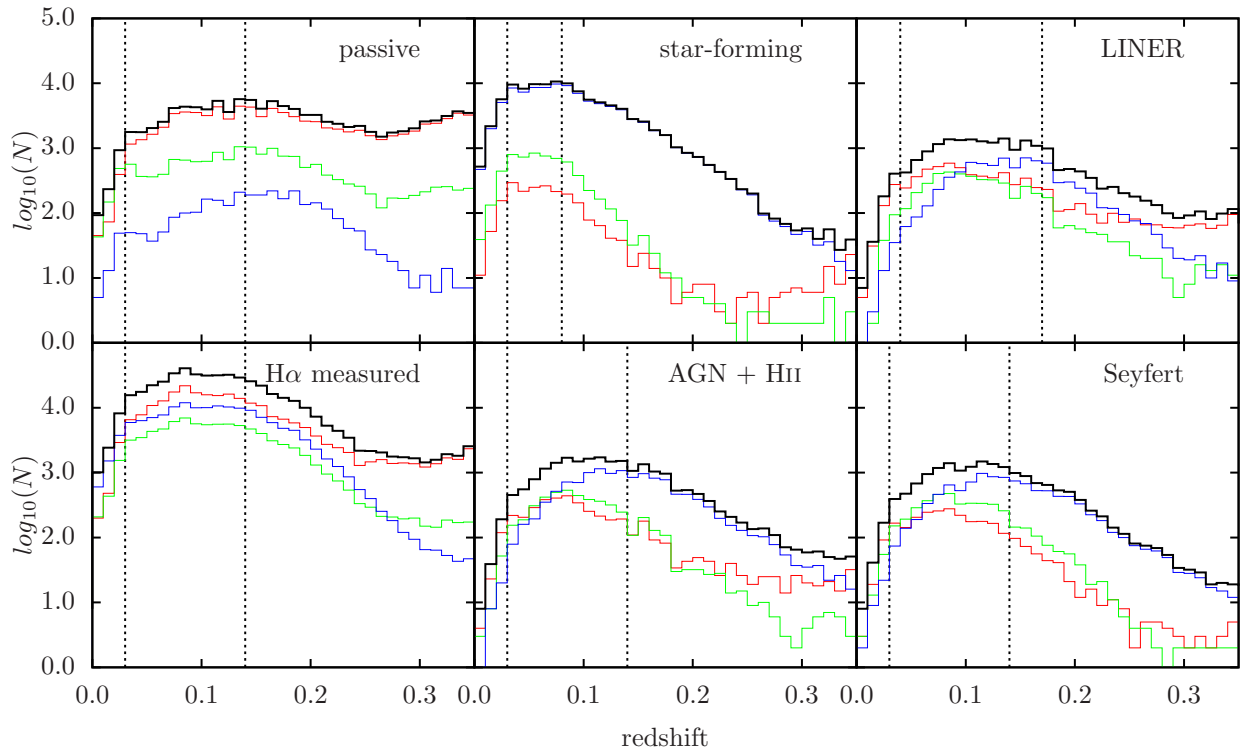}
      \vspace{36pt}
      \caption{Redshift distribution of the galaxies in each activity class \textit{before} the absolute magnitude cuts. Colours represent the colour classes: red -- red galaxies, green -- green valley galaxies, blue -- blue galaxies, and black -- all galaxies regardless of their colours. The dashed vertical lines represent the redshift cuts, which are different for every activity class.}
      \label{fig:redshiftdist}
    \end{center}
   \end{minipage}
  \end{figure*}

\begin{figure*}
   \begin{minipage}{1\textwidth}
    \begin{center}
	  \input{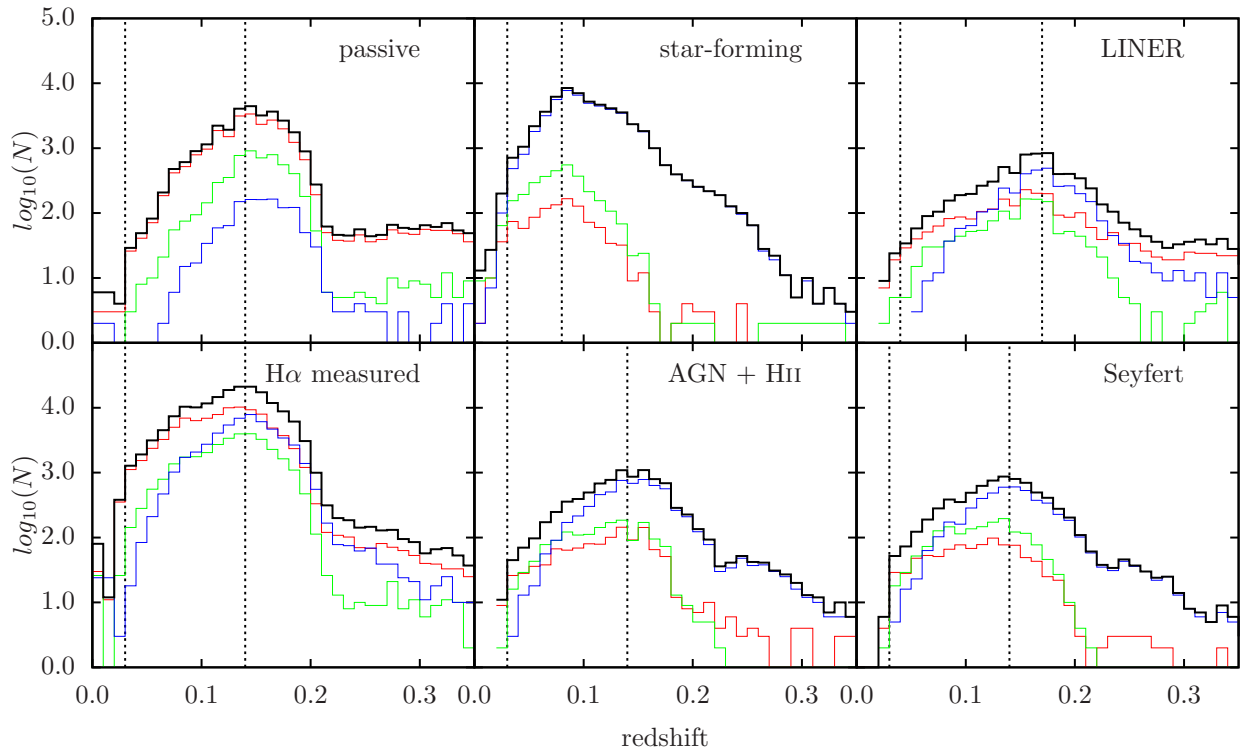}
      \vspace{36pt}
      \caption{Redshift distribution of the galaxies in each activity class \textit{after} the absolute magnitude cuts. Colours represent the colour classes: red -- red galaxies, green -- green valley galaxies, blue -- blue galaxies, and black -- all galaxies regardless of their colours. The dashed vertical lines represent the redshift cuts, which are different for each activity class.}
      \label{fig:redshiftdist_cut}
    \end{center}
   \end{minipage}
  \end{figure*}

The redshift distribution of the galaxies is plotted in Fig.~\ref{fig:redshiftdist} for every galaxy activity class in redshift bins of $\Delta z = 0.01$. It is clear from the curves that all samples except red galaxies are flux-limited according to the spectroscopic targeting algorithm of SDSS in the investigated redshift range. The bright-end red galaxy sample is volume-limited to $z = 0.35$ \citep{Eisenstein2001}. Fig.~\ref{fig:redshiftdist_cut} shows the same redshift distributions \textit{after} applying the absolute magnitude cuts to get the volume-limited samples. Note that the requirement to select volume limited samples does not allow maximizing galaxy counts by freely choosing the used redshift intervals; absolute magnitude limits also have to be taken into account.

\section{Composite spectra}
\label{sec:composite}

\subsection{Preprocessing}

We selected spectra based on the criteria defined in Sec.~\ref{sec:class} and preprocessed them before calculating the composites. First, all spectra were corrected for foreground galactic extinction using the dust map of \citet{Schlegel1998} and the extinction curve from \citet{ODonnell1994}. Second, spectra were normalised as described in Sec.~\ref{sec:norm}. Third, spectra were converted to rest-frame and re-binned to a grid with the intended resolution ({$\Delta \lambda = 1.0$, $ 0.5 $ or $ 0.25 $ \AA}).

\subsection{Normalisation}
\label{sec:norm}

To derive meaningful composites, one has to normalise the galaxy spectra in a robust way. Normalising spectra at a given wavelength is not appropriate for noisy measurements, thus we use the following technique. We calculate the median flux in the $4200 - 4300$ \AA, $4600-4800$ \AA, $5400-5500$ \AA, and $5600-5800$ \AA \ wavelength intervals and use this value to scale spectra together. The wavelength intervals are chosen to be devoid of any strong emission lines. Spectra are re-scaled to have the median value of the flux in these intervals equal to one.

We do not weight or otherwise scale the flux of the spectra based on the apparent or intrinsic luminosity of the objects or S/N of the observations. Although this could affect the signal to noise ratio of the composites negatively, the robust algorithm can successfully overcome this problem by automatically assigning lower weights to spectra with lower signal to noise values.

\subsection{Robust algorithm}

In order to take the significant noise and the varying signal to noise ratio of the original data set into account, we use the robust averaging method of \citet{Budavari2009} to determine the average and variance spectra of the different galaxy classes. This algorithm not only computes the average and the variance iteratively by taking care of gaps in the measurements, but also yields the first few principal components of the spectrum distribution (see Sec.~\ref{sec:pca}). We will detail the efficiency of the robust method with respect to ordinary average and median composite spectra in Sec.~\ref{sec:perf}.

  \begin{figure*}
   \begin{turn}{90}
    \begin{minipage}{1\textheight}
     \begin{center}
      \input{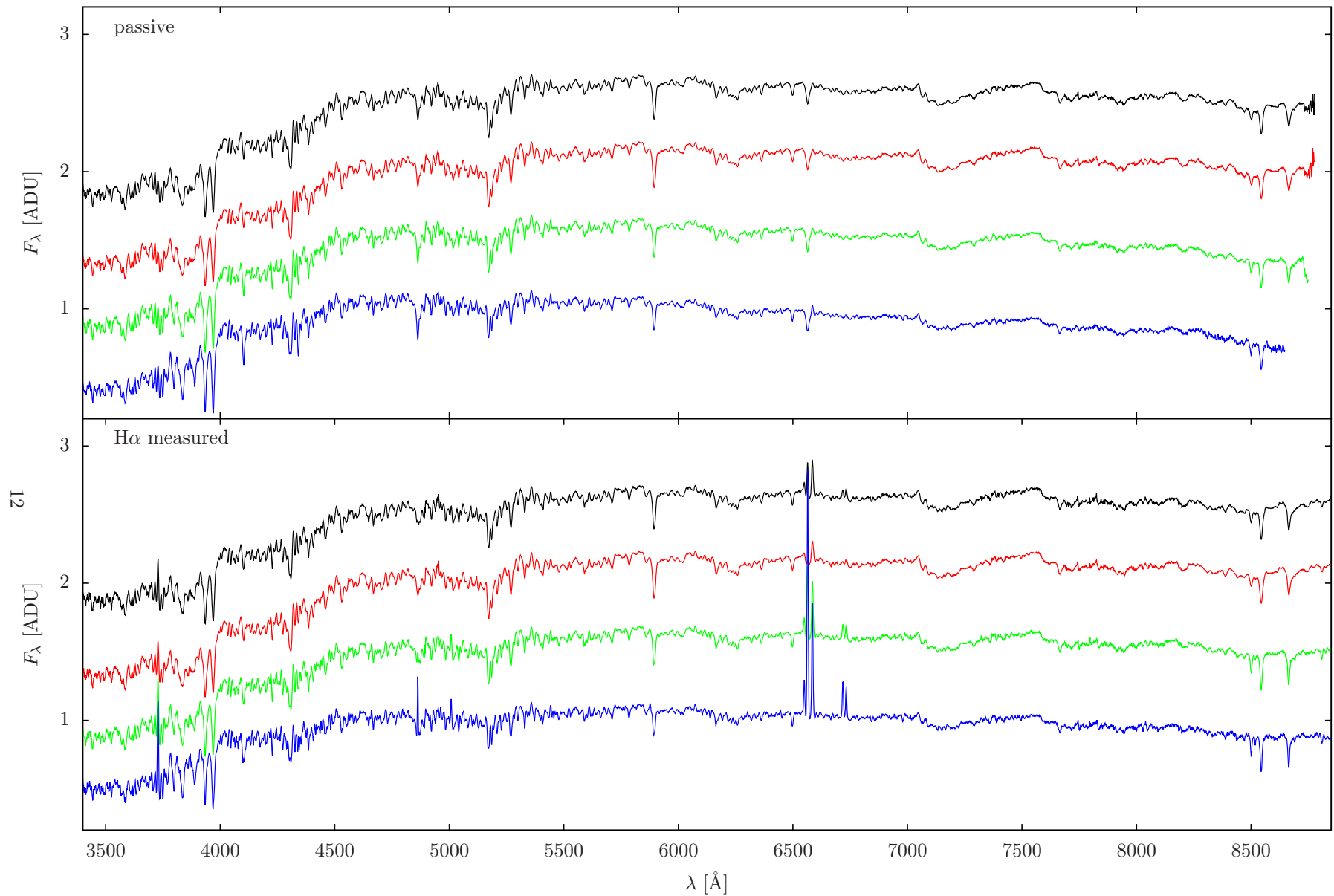}
      \vspace{32pt}
      \caption{Composite spectra of the different activity and colour classes. Activity class is indicated in the top left corner of each panel, the spectra of the different colour classes are offset by $ 0.5 $ along the $ y $ axis for clarity, and indicated by colours from top to bottom (black: all colour, red, green and blue). See the online edition for a colour version of this plot.}
      \label{fig:composites1}      
     \end{center}
    \end{minipage}
   \end{turn}
  \end{figure*}
  
  \begin{figure*}
   \begin{turn}{90}
    \begin{minipage}{1\textheight}
     \begin{center}
      \label{fig:composites2}
      \input{figures/composites2.tex}
      \vspace{32pt}
      \caption{See caption of Fig.~\ref{fig:composites1} for detailed description.}
     \end{center}
    \end{minipage}
   \end{turn}
  \end{figure*}
  
  \begin{figure*}
   \begin{turn}{90}
    \begin{minipage}{1\textheight}
     \begin{center}
      \label{fig:composites3}
      \input{figures/composites3.tex}
      \vspace{32pt}
      \caption{See caption of Fig.~\ref{fig:composites1} for detailed description.}
     \end{center}
    \end{minipage}
   \end{turn}
  \end{figure*}     
  
  \begin{figure*}
   \begin{turn}{90}
    \begin{minipage}{1\textheight}
     \begin{center}
      \input{figures/composites4.tex}
      \vspace{32pt}
      \caption{Composite spectra of the refined colour classes as described in Sec.~\ref{sec:classrefine}. The curves are colour-coded from blue (top) to red (bottom) based on the $ g - r $ colour of the galaxies. See the online edition for a colour version of this plot.}
      \label{fig:composites4}   
     \end{center}
    \end{minipage}
   \end{turn}
  \end{figure*}     
  
\textcolor{black}{The composites calculated by the robust algorithm are plotted in Figs.~\ref{fig:composites1}--\ref{fig:composites4} for each activity and colour class.}

\subsection{High resolution}

Because composites are combined from many spectra at varying redshifts, it is possible to get a couple of factors higher resolution than the resolution of the original measurements. The average error in redshift measurement of the high signal-to-noise ($S/N \geq  30$) galaxy spectra of SDSS is $z_{\text{err}} = 1.6 \times 10^{-4}$, which allows room for oversampling the composites. On the other hand, oversampling can introduce noise compared to the almost noiseless composites at 1 \AA \ binning. This noise can become significant for sparsely populated galaxy classes such as the green valley galaxies.

Another important factor that counteracts the efforts to get higher resolution is the intrinsic broadening of the spectral features of the individual spectra due to the high velocity kinetics in the galaxies. To overcome this problem, deconvolution of the Doppler broadening kernel from the original measurements would be necessary. We do not address this issue in the current work.

\begin{figure}
    \begin{center}
      \input{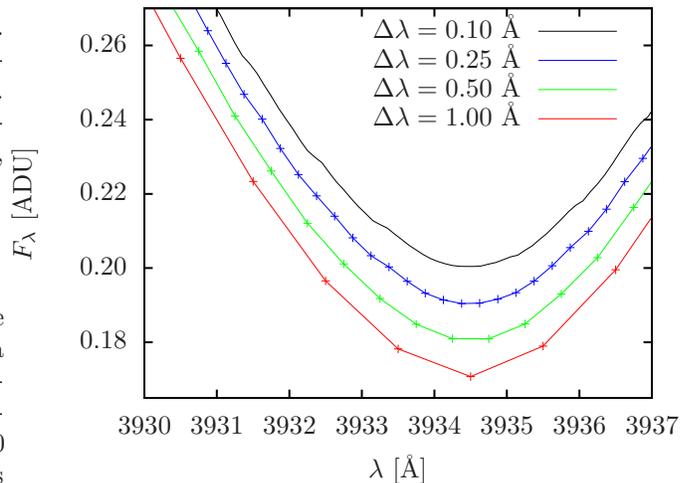}
      \caption{The K line of a galaxy composite spectrum calculated at resolutions $ \Delta\lambda = 1, 0.5, 0.25 $ and $ 0.1 $ \AA (from bottom to top). Higher resolution curves are smoother compared to the {$\Delta\lambda = 1$ \AA} curve.}
      \label{fig:resolution}
    \end{center}
  \end{figure}

Fig.~\ref{fig:resolution} shows the comparison of some spectral features of a galaxy composite with different oversampling (at the resolution of 1 \AA, 0.5 \AA, 0.25 \AA, and 0.1 \AA). Note how the line features become smoother with higher resolution. It is important, though, that oversampled spectra might not contain more information because the wide velocity dispersion washes out fine details.

Evidently, linear interpolation could not be used to get the higher resolution composites form lower resolution ones. Higher-order curves could be used for oversampling low resolution composites to get smooth line profiles, but they are likely to introduce artefacts. Our method of oversampling overcomes this problem.

\subsection{Generating empirical mock catalogues using eigenspectra}
\label{sec:pca}

Besides the robust average spectrum of every galaxy class, we also calculate the variance spectra. These can be used to determine the wavelength ranges with outstanding variance (usually the emission lines) and quantify the noise in each wavelength bin. The variance spectra are available on-line. However, if one is interested in the overall variance of the spectra (in contrast to the bin-by-bin variance) of a galaxy class, it is not enough to use the variance spectrum where bins are treated independently. In order to quantify the variance among the spectra of a certain galaxy class over the entire wavelength range, and take the covariance of the bins into account, the first few principal components (eigenspectra) have to be calculated. Fortunately, our robust averaging algorithm yields the eigenspectra as a ``by-product'' of the averaging.

We calculate the first five eigenspectra of every galaxy class, and expand each measured spectrum on this basis. In most cases, the distribution of the expansion coefficients is very close to a Gaussian distribution centred on zero. We determine the standard deviation $\sigma_{\text{eig}}$ of the distribution of the eigencoefficients in each galaxy class. Then one can calculate the following spectra.

  \begin{equation}
	F_{ \lambda }^{\pm } = F_{ \lambda }^{\text{avg} } \pm k \sigma_{\text{eig}} F_{ \lambda }^{\text{eig}},
	\label{eq:varspectra}
  \end{equation}
  
where $F_{ \lambda }^{\text{avg} }$ and $F_{ \lambda }^{\text{eig} }$ are the average and one of the eigenspectra, respectively. The coefficient $ k $ is a constant factor that can be varied to get slightly different spectra.

\begin{figure*}
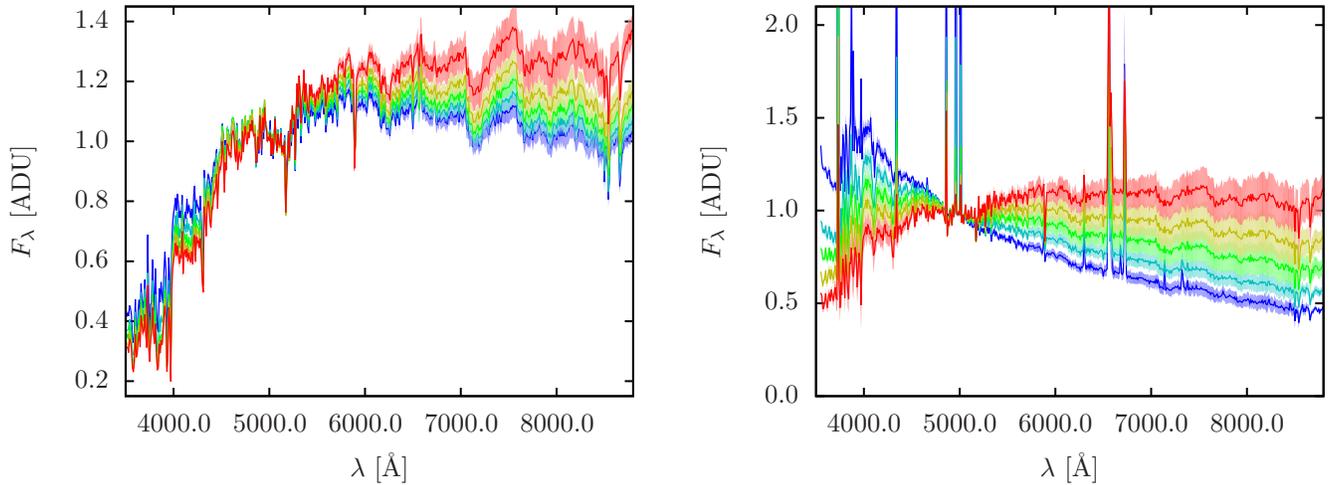

      \begin{minipage}{1\textwidth}
    \begin{center}
		\subfigure{\input{figures/var_passive.tex}}
		\hspace{16pt}
    	\subfigure{\input{figures/var_star.tex}}
      \caption{Variance in red (left panel) and blue star-forming (right panel) galaxies. The solid curves show the composites of galaxies classified according to the refined colour classes (redder to bluer). The shaded areas represent the $0.5 \sigma$ (red galaxies) and $ 1 \sigma $ (blue galaxies) variations about the average. See the online edition for a colour version of this plot. }
      \label{fig:var}
    \end{center}
    \end{minipage}
  \end{figure*}

As an illustration, we plot $F_{ \lambda }^{\pm }$ for the refined classes of red and blue galaxies in Fig.~\ref{fig:var}. In case of the red galaxies, the first eigenspectrum clearly determines the colour of the galaxy; almost the entire variation in the overall slope of spectrum is picked up by the first principal component. In case of the star-forming galaxies, most of the variance is in the emission lines, so one has to use the second eigenvector to generate spectra with colour variations. As Fig.~\ref{fig:var} shows, this technique can be successfully used to generate composites of any colour between the averages of the predefined classes. One drawback of the method is that adding higher order eigenspectra to the averages introduces some noise.

\textcolor{black}{The importance of this technique is that it allows generating mock spectral libraries. By drawing the value of $k$ of Eq.~\ref{eq:varspectra} from a normal distribution, the $F_{ \lambda }^{\pm }$ spectra will be representative to the whole ensemble of SDSS galaxies. Eigenspectra of the various galaxy classes are published online along with the parameters (mean and $\sigma$) of the distributions of the expansion coefficients of individual observations on the eigenbases. By using these parameters, linear combinations of the average composite spectra and eigenspectra can be easily calculated to further extend the parameter space coverage of the presented atlas.}
  
\subsection{Performance of the robust method}
\label{sec:perf}

\textcolor{black}{To quantify the performance of the robust method, we compare the robustly averaged composites with normally averaged and median composites. Averaging is thought to result in composites that are physical in the sense that they are constructed as non-negative linear combinations of real stellar continua. Median, while it is robust against noise, differs from averaging because it treats the spectral bins independently and might yield non-physical continua. One disadvantage of averaging spectra it that emission line ratios are not guaranteed to be preserved, whereas \citet{VandenBerk2001} reported that taking the median spectra of quasars did not alter the emission line ratios. We will revisit the question of emission lines of average composites in Sec.~\ref{sec:emlines}. The main advantage of robust averaging over median calculation is that the former can be done iteratively, and there is no need to store and sort all the data in memory.}

\begin{figure*}
      \begin{minipage}{1\textwidth}
    \begin{center}
      \hspace{32pt} \input{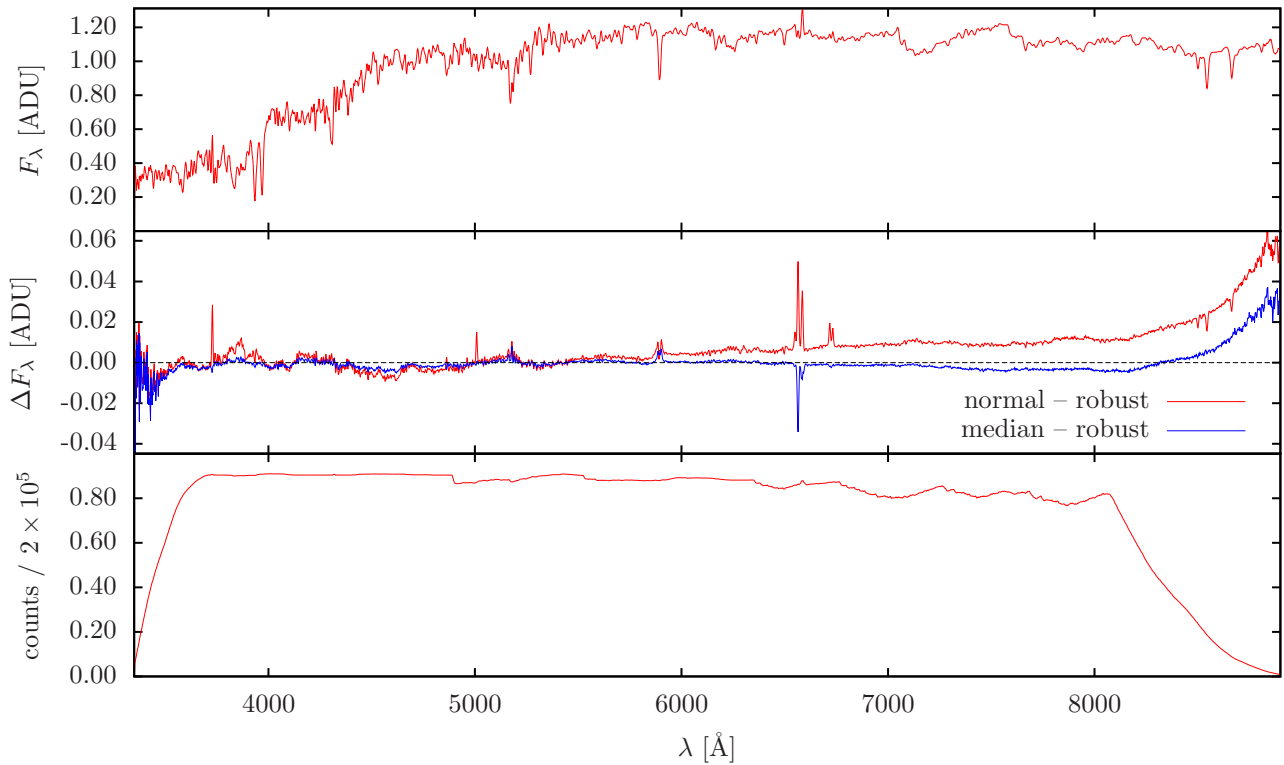}
      \vspace{32pt}
      \caption{Comparision of the various averaging methods for red galaxies with measured H$ \alpha $. We plot the spectrum in the top panel for reference. The middle panel shows the difference between the robust average and the normal average and the median, respectively. The bottom panel shows the number of spectra used to determine the averages of the wavelength bins.}
      \label{fig:robust_red}
    \end{center}
    \end{minipage}
  \end{figure*}

\begin{figure*}
      \begin{minipage}{1\textwidth}
    \begin{center}
      \hspace{32pt} \input{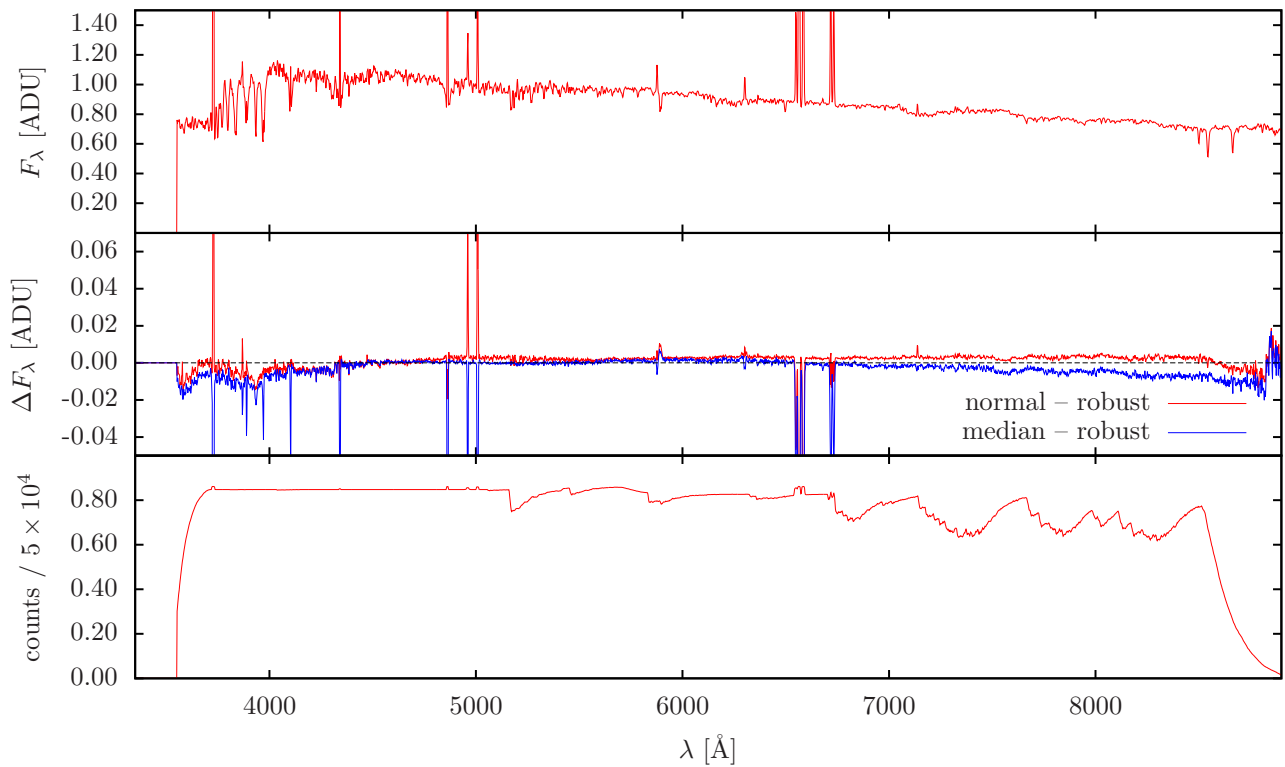}
      \vspace{32pt}
      \caption{Same as Fig.~\ref{fig:robust_red}, but for star-forming blue galaxies.}
      \label{fig:robust_blue}
    \end{center}
    \end{minipage}
  \end{figure*}

We plot the difference between the robust average and the non-robust composites in Fig.~\ref{fig:robust_red} and \ref{fig:robust_blue} (middle panels). The robust composite and the number counts used to calculate the composites in each wavelength bin are also plotted for reference in the top and bottom panels, respectively. 

Interestingly, for red galaxies, the median runs much closer to the robust average than the normal average, while in case of blue galaxies the differences are not as significant as in case of the red galaxies. Significant differences between the robust and non-robust composites are visible at very short and very long wavelengths, and also at emission lines. The differences between the continua are possibly due to the low number counts in those wavelength ranges. For low number counts, non-robust methods tend to become severely biased. \textcolor{black}{The robust algorithm assigns lower weights to galaxies with extremely strong emission lines, which is considered to be the reason behind the large difference between the robust and non-robust methods in the emission line regions.}

\section{Composites on the map}
\label{sec:map}

\textcolor{black}{In this section, we use standard techniques to estimate the stellar continua, emission line strengths, absorption line indices and physical parameters of the composites. We follow the methods described by \citet{Kauffmann2003a, Kauffmann2003b, Brinchmann2004} and \citet{Tremonti2004}. These techniques were developed and used by the the cited authors to compose the MPA/JHU value-added galaxy catalogue\footnote{\url{http://www.mpa-garching.mpg.de/SDSS/}} of SDSS DR7.}

\textcolor{black}{In this paper, we do not intend to make any conclusions about physical parameters of the galaxy samples based on the parameters derived from the composites. Our goal is simply to show that the composites calculated using our method cover much of the parameter space of the individual galaxies. To achieve this, we plot several standard diagrams widely used in the literature. These diagrams show the distributions of the parameters of individual galaxies of the SDSS or MPA/JHU value-added catalogues overplotted with the parameters of the composites.}

\subsection{Continuum model fits}
\label{sec:models}

The wavelength coverage of our empirically-derived high signal to noise composite spectra spans from about $3350$ \AA \  to $9050$ \AA. The actual coverage depends on the redshift cuts of the volume-limited samples of the galaxy classes; higher redshift samples allow extending the coverage blueward. In order to extend this limited coverage, we fit the composites with theoretical model spectra generated using the single stellar population (SSP) synthesis models of \citet{BC03}. Three sets of models were generated with different metallicity values of $Z = 0.008$, $0.02$, and $0.05$. Each set contains 10 single burst models with ages of $t = 0.005$, $0.025$, $0.100$, $0.290$, $0.640$, $0.900$, $1.4$, $2.5$, $5.0$ and $11.0$ Gyr based on the initial mass function (IMF) of \citet{Chabrier2003} and the evolutionary tracks `Padova1994' \citep{Alongi1993, Bressan1993, Fagotto1994a, Fagotto1994b}. We use the non-negative least squares (NNLS) method to determine the best linear combination of models of different age (but same metallicity) to fit the composite spectra. To estimate the effect of intrinsic extinction by dust, we use the simple dust model of \citet{BC03}, which is described by the following formula:

\begin{equation*}
	\tau_{\lambda}(t) = \left \{
		\begin{array}{ r c l }
		\tau_{V}(\lambda / 5500 \: \text{\AA})^{-0.7} & \text{for} & t \leq 10^{7} \text{yr} \\
		\mu \tau_{V}(\lambda / 5500 \: \text{\AA})^{-0.7} & \text{for} & t > 10^{7} \text{yr}
		\end{array}
		\right .
\end{equation*}

We let the value of the optical depth $\tau_V$ vary while keeping the value of $\mu = 0.3$ fixed. In this particular dust model, $\mu$ controls the ratio of the effect of extinction on stars older/younger than $10$ Myr. We also keep the value of the velocity dispersion of the stars $\sigma_v$ a free parameter. For each composite, we repeat the fitting for each model set, and the metallicity of the best fitting model set is used.

To compute the synthetic magnitude of the composites in broadband filters where the wavelength coverage of the composites is not wide enough, we extend the empirical spectra with the best fitting models. We also use the same continuum fits to measure the emission line equivalent widths of the composites.

\subsection{Colour-colour trajectories}

\begin{figure*}
  \begin{minipage}{1\textwidth}
    \begin{center}
	  \vspace{24pt}
	  \hspace{18pt}
      \input{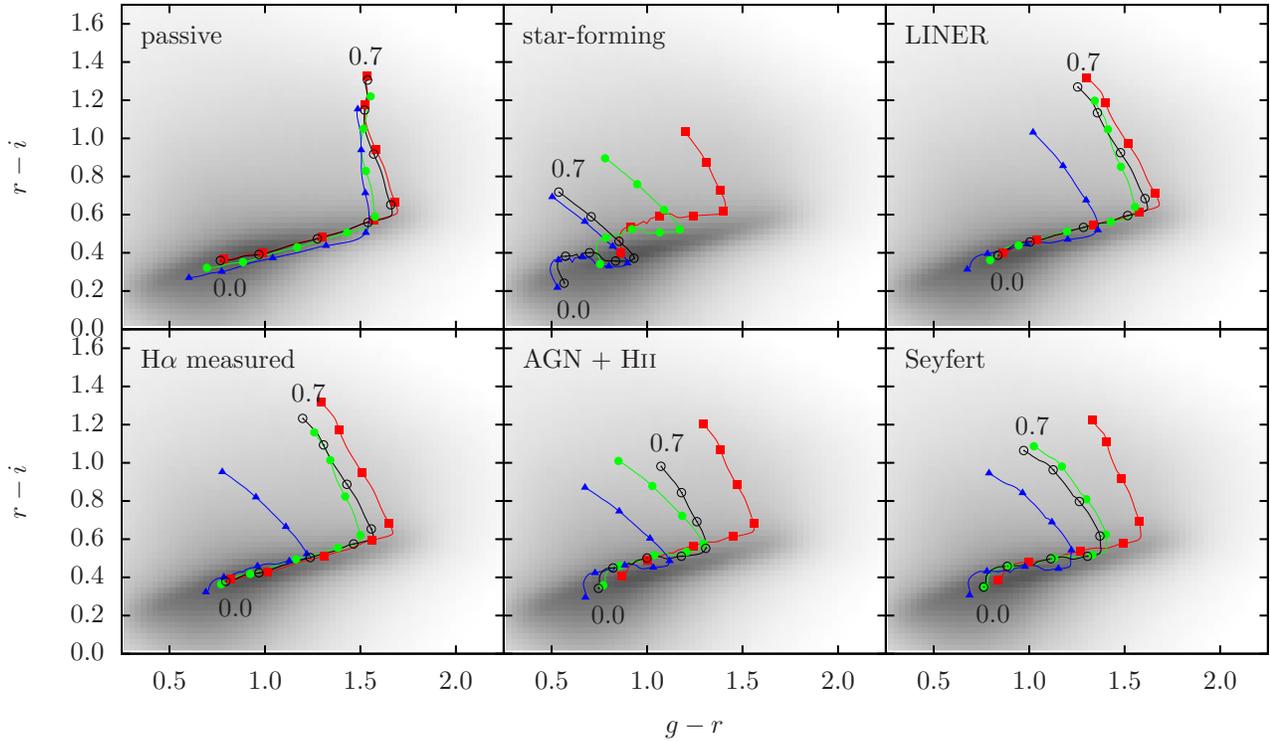}
      \vspace{36pt}
      \caption{Colour-colour diagram of the composites for the SDSS $g - r$ and $r - i$ colour indices between redshifts $0 \leq z \leq 0.7$. The coloured curves represent the composite spectra of the different colour classes we defined in Sec.~\ref{sec:colorcut}. The black curve represents the composite spectra calculated from galaxies with any colour. Dots along the curves mark redshifts $z = 0.0, 0.1$, etc. The greyscale map in the background shows the distribution of the extinction corrected photometric colours calculated from the fibre magnitudes of SDSS \texttt{photo} galaxies, which is significantly deeper than the \texttt{spectro} set.}
      \label{fig:traject_gri}
    \end{center}
    \end{minipage}
  \end{figure*}

\begin{figure}
  \begin{minipage}{1\columnwidth}  
    \begin{center}
      \hspace{24pt}\input{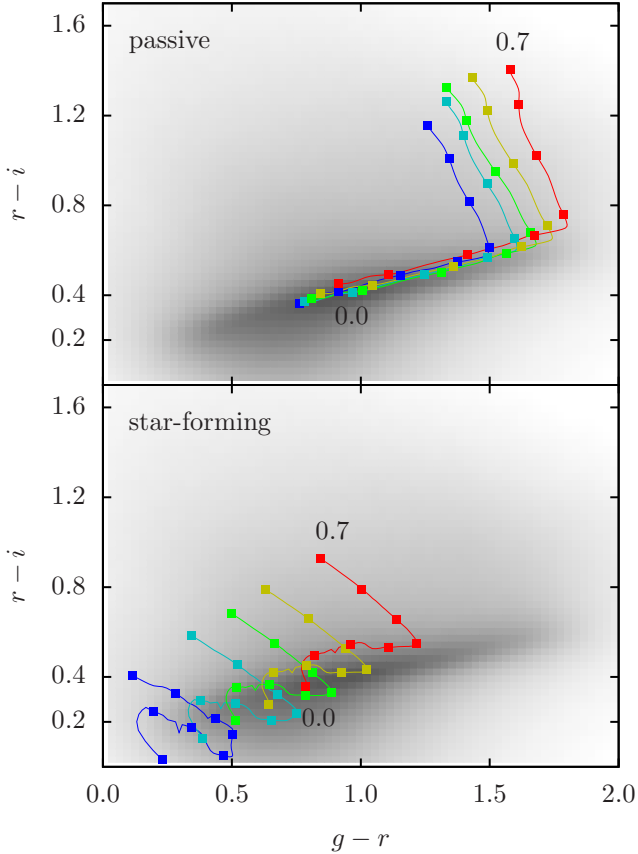}
      \vspace{36pt}
      \caption{Colour-colour diagrams of the composites for the SDSS $g - r$ and $r - i$ colour indices between redshifts $0 \leq z \leq 0.7$. The coloured curves represent the composite spectrum of the different refined colour classes (top panel: red galaxies, bottom panel: star-forming galaxies), as defined in Sec.~\ref{sec:classrefine}. Dots along the curves mark redshifts $z = 0.0, 0.1$, etc. The greyscale map in the background shows the distribution of the photometric colours of SDSS \texttt{spectro} galaxies.}
      \label{fig:traject_gri_refined}
    \end{center}
    \end{minipage}    
  \end{figure}

\textcolor{black}{As we intend to use the composite atlas as a basis for template-based photometric redshift estimation, it is important that the composites cover much of the colour space spanned by the SDSS galaxies.} Fig.~\ref{fig:traject_gri} and \ref{fig:traject_gri_refined} display the SDSS $g-r$; $r-i$ colour-colour trajectories of composites for every galaxy class we defined. The trajectories cover the redshift range of $0.0 \leq z \leq 0.7$, and they nicely overlap with the measured galaxy distributions. This is a good confirmation that the robust averaging algorithm does not introduce any bias into galaxy colours.

\textcolor{black}{It is observable in Fig.~\ref{fig:traject_gri} that the bluest galaxies ($g - r < 0.5$) are not covered by the composites of the main galaxy activity classes. The refined classification of star-forming galaxies however yields composites of extremely blue colours, as it can be seen in \ref{fig:traject_gri_refined}.}

\subsection{Emission line measurements}
\label{sec:emlines}

\textcolor{black}{As averaging spectra may have the effect of altering line ratios, we have to verify whether the emission lines of the composites are physical or not. Fortunately, in our case the robust averaging algorithm does not alter the line ratios severely, as we will show below.}

\begin{figure}
    \begin{center}
      \input{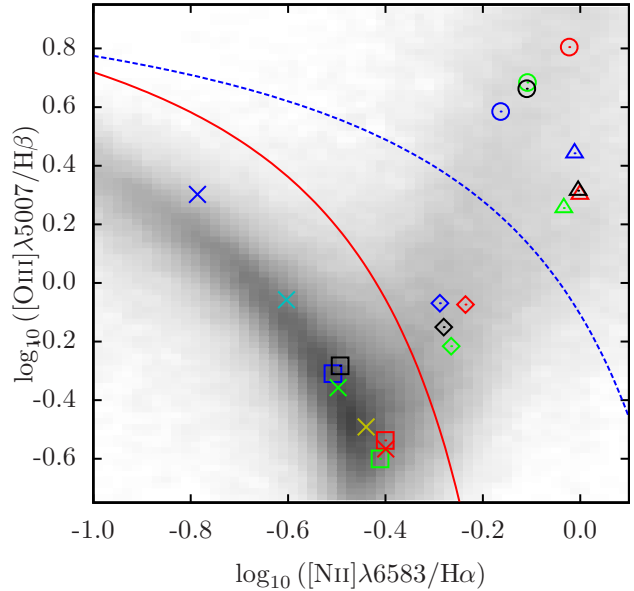}
      \vspace{10pt}
      \caption{BPT diagram of the composite spectra. The density plot is the logarithm of the number density of all SDSS galaxies. The solid red curve indicates the segregation line between star forming galaxies and active galactic nuclei. The blue dashed curve is the extreme starburst line. The loci of the composites are indicated with points: rectangles -- star-forming galaxies; diamonds -- AGN+\ion{H}{ii} galaxies, triangles -- LINERs, circles -- Seyferts. The different colours denote the different colour classes; black points mark the loci of composites of all galaxies (regardless of their colours) in a given activity class. Colour crosses mark the loci of the star-forming galaxies, according to the refined classification. See the online edition for a colour version of this plot.}	
      \label{fig:bptcomposites}
    \end{center}
  \end{figure}

\textcolor{black}{First, emission line equivalent widths are determined by using the best fit continuum model as described in Sec.~\ref{sec:models}. Then,} the BPT diagram is plotted for the composites of the star forming, AGN + \ion{H}{ii}, LINER and Seyfert galaxies in Fig.~\ref{fig:bptcomposites}. The loci of the composites overlap with the line ratios calculated from the individual galaxy measurements. It is important to mention, however, that the variance in line ratios of galaxies is much larger than what a few composite spectra can characterize.

One interesting feature of Fig.~\ref{fig:bptcomposites} is that Seyferts are located along a well-defined line parametrized by the $ g-r $ colour: redder Seyferts tend to have higher [\ion{O}{iii}]/H$\beta$ ratios. This is consistent with the findings of \citet{Kewley2006}: the stellar populations of AGNs are the older the further away they are located from the star-forming line on the BPT diagram. For AGN + \ion{H}{ii} galaxies, no similar behaviour is readily visible.

In case of the star-forming galaxies, we also plot the results from the refined colour classification. A clear trend is observable here too: bluer galaxies have higher [\ion{O}{iii}]/H$\beta$ ratios and lower [\ion{N}{ii}]/H$\alpha$ ratios. \textcolor{black}{Bluer star-forming composites (crosses), however, tend to be slightly above the star-forming line of the MPA/JHU results (density plot). The most trivial explanation to this effect is that averaging alters the line ratios and this bias is more significant in case of stronger emission lines.}

Emission line equivalent widths of all composites are listed in Tab.~\ref{tab:lines1} -- \ref{tab:lines3} and are available on-line.

\begin{table*}
   \begin{turn}{90}
   \begin{minipage}{1\textheight}
		\begin{center}
			\begin{tabular}{ | l | r ||  r | r | r | r | r | r | r || r | r | r | r | r | r | r | }
\hline
 & $\lambda$ & red & red & red & red & red & red & red & green & green & green & green & green & green & green \\
 & [\AA]  & P & H$\alpha$ & SF & A+\ion{H}{ii} & L & S & all & P & H$\alpha$ & SF & A+\ion{H}{ii} & L & S & all \\
 \hline
 \hline
\ion{O}{ii} & 3727.09 & - & 4.163 & 14.122 & 7.523 & 7.337 & 8.480 & 5.202 & - & 2.369 & 13.465 & 7.680 & 9.742 & 9.010 & 2.847 \\
\ion{O}{ii} & 3729.88 & - & - & 4.857 & 8.665 & 7.729 & 9.871 & - & - & 1.467 & 17.105 & 8.734 & 10.975 & 10.246 & 1.723 \\
H$\theta$ & 3798.98 & - & - & - & - & - & - & - & - & - & - & - & - & - & - \\
H$\eta$ & 3836.47 & - & - & - & - & - & - & - & - & - & - & - & - & - & - \\
\ion{He}{i} & 3889.00 & - & - & 1.390 & 1.111 & - & 1.261 & - & - & - & 1.382 & - & - & 1.197 & - \\
\ion{S}{ii} & 4072.30 & - & - & - & - & - & - & - & - & - & - & - & - & - & - \\
H$\delta$ & 4102.89 & - & - & 2.068 & 1.235 & - & 1.131 & - & - & - & 2.029 & 1.252 & - & - & - \\
H$\gamma$ & 4341.68 & - & 1.433 & 3.889 & 2.464 & - & 2.235 & 1.868 & - & - & 3.766 & 1.888 & - & 1.730 & - \\
\ion{O}{iii} & 4364.44 & - & - & - & - & - & - & - & - & - & - & - & - & - & - \\
H$\beta$ & 4862.68 & - & 3.091 & 9.453 & 5.607 & 1.471 & 4.784 & 4.087 & - & - & 9.060 & 4.745 & 1.159 & 3.401 & - \\
\ion{O}{iii} & 4932.60 & - & - & - & - & - & - & - & - & - & - & - & - & - & - \\
\ion{O}{iii} & 4960.30 & - & - & 1.448 & 1.121 & - & 5.680 & - & - & - & 1.308 & - & - & 5.063 & - \\
\ion{O}{iii} & 5008.24 & - & 1.086 & 5.125 & 4.198 & 2.949 & 18.697 & 1.998 & - & - & 4.692 & 3.252 & 2.622 & 16.649 & - \\
\ion{O}{i} & 6302.05 & - & - & 1.272 & 1.228 & 1.580 & 1.611 & - & - & - & 1.247 & 1.238 & 1.490 & 1.496 & - \\
\ion{O}{i} & 6365.54 & - & - & - & - & - & - & - & - & - & - & - & - & - & - \\
\ion{N}{i} & 6529.03 & - & - & - & - & - & - & - & - & - & - & - & - & - & - \\
\ion{N}{ii} & 6549.86 & - & - & 4.204 & 5.429 & 5.731 & 5.875 & 3.902 & - & - & 4.150 & 6.879 & 3.603 & 7.439 & 1.658 \\
H$\alpha$ & 6564.61 & - & 14.908 & 41.231 & 26.957 & 10.188 & 22.476 & 19.225 & - & 3.751 & 39.508 & 21.372 & 6.435 & 16.527 & 4.380 \\
\ion{N}{ii} & 6585.27 & - & 5.677 & 12.908 & 13.600 & 9.450 & 15.528 & 7.629 & - & 2.137 & 12.745 & 11.214 & 6.453 & 12.834 & 2.436 \\
\ion{S}{ii} & 6718.29 & - & 2.053 & 7.254 & 4.735 & 3.932 & 4.790 & 2.853 & - & - & 6.906 & 4.380 & 3.617 & 4.245 & 1.100 \\
\ion{S}{ii} & 6732.67 & - & 1.478 & 5.205 & 3.748 & 3.351 & 4.242 & 2.088 & - & - & 4.973 & 3.682 & 3.316 & 3.932 & - \\
\hline
\end{tabular}

		\end{center} 
		 \caption{Values of the emission line equivalent widths (in \AA ngstr\"oms) for every galaxy class, continued in Tab.~\ref{tab:lines2}.}
 		\label{tab:lines1}
 	\end{minipage}
 	\end{turn} 	
\end{table*}

\begin{table*}
   \begin{turn}{90}
   \begin{minipage}{1\textheight}
		\begin{center}
			\begin{tabular}{ | l | r ||  r | r | r | r | r | r | r || r | r | r | r | r | r | r | }
\hline
 & $\lambda$ & blue & blue & blue & blue & blue & blue & blue & all & all & all & all & all & all & all \\
 & [\AA]  & P & H$\alpha$ & SF & A+\ion{H}{ii} & L & S & all & P & H$\alpha$ & SF & A+\ion{H}{ii} & L & S & all \\
 \hline
 \hline
\ion{O}{ii} & 3727.09 & - & 4.163 & 14.122 & 7.523 & 7.337 & 8.480 & 5.202 & - & 2.369 & 13.465 & 7.680 & 9.742 & 9.010 & 2.847 \\
\ion{O}{ii} & 3729.88 & - & - & 4.857 & 8.665 & 7.729 & 9.871 & - & - & 1.467 & 17.105 & 8.734 & 10.975 & 10.246 & 1.723 \\
H$\theta$ & 3798.98 & - & - & - & - & - & - & - & - & - & - & - & - & - & - \\
H$\eta$ & 3836.47 & - & - & - & - & - & - & - & - & - & - & - & - & - & - \\
\ion{He}{i} & 3889.00 & - & - & 1.390 & 1.111 & - & 1.261 & - & - & - & 1.382 & - & - & 1.197 & - \\
\ion{S}{ii} & 4072.30 & - & - & - & - & - & - & - & - & - & - & - & - & - & - \\
H$\delta$ & 4102.89 & - & - & 2.068 & 1.235 & - & 1.131 & - & - & - & 2.029 & 1.252 & - & - & - \\
H$\gamma$ & 4341.68 & - & 1.433 & 3.889 & 2.464 & - & 2.235 & 1.868 & - & - & 3.766 & 1.888 & - & 1.730 & - \\
\ion{O}{iii} & 4364.44 & - & - & - & - & - & - & - & - & - & - & - & - & - & - \\
H$\beta$ & 4862.68 & - & 3.091 & 9.453 & 5.607 & 1.471 & 4.784 & 4.087 & - & - & 9.060 & 4.745 & 1.159 & 3.401 & - \\
\ion{O}{iii} & 4932.60 & - & - & - & - & - & - & - & - & - & - & - & - & - & - \\
\ion{O}{iii} & 4960.30 & - & - & 1.448 & 1.121 & - & 5.680 & - & - & - & 1.308 & - & - & 5.063 & - \\
\ion{O}{iii} & 5008.24 & - & 1.086 & 5.125 & 4.198 & 2.949 & 18.697 & 1.998 & - & - & 4.692 & 3.252 & 2.622 & 16.649 & - \\
\ion{O}{i} & 6302.05 & - & - & 1.272 & 1.228 & 1.580 & 1.611 & - & - & - & 1.247 & 1.238 & 1.490 & 1.496 & - \\
\ion{O}{i} & 6365.54 & - & - & - & - & - & - & - & - & - & - & - & - & - & - \\
\ion{N}{i} & 6529.03 & - & - & - & - & - & - & - & - & - & - & - & - & - & - \\
\ion{N}{ii} & 6549.86 & - & - & 4.204 & 5.429 & 5.731 & 5.875 & 3.902 & - & - & 4.150 & 6.879 & 3.603 & 7.439 & 1.658 \\
H$\alpha$ & 6564.61 & - & 14.908 & 41.231 & 26.957 & 10.188 & 22.476 & 19.225 & - & 3.751 & 39.508 & 21.372 & 6.435 & 16.527 & 4.380 \\
\ion{N}{ii} & 6585.27 & - & 5.677 & 12.908 & 13.600 & 9.450 & 15.528 & 7.629 & - & 2.137 & 12.745 & 11.214 & 6.453 & 12.834 & 2.436 \\
\ion{S}{ii} & 6718.29 & - & 2.053 & 7.254 & 4.735 & 3.932 & 4.790 & 2.853 & - & - & 6.906 & 4.380 & 3.617 & 4.245 & 1.100 \\
\ion{S}{ii} & 6732.67 & - & 1.478 & 5.205 & 3.748 & 3.351 & 4.242 & 2.088 & - & - & 4.973 & 3.682 & 3.316 & 3.932 & - \\
\hline
\end{tabular}

		\end{center} 
		 \caption{Values of the emission line equivalent widths (in \AA ngstr\"oms) for every galaxy class, continued from Tab.~\ref{tab:lines1}.}
 		\label{tab:lines2}
 	\end{minipage}
 	\end{turn} 	
\end{table*}

\begin{table*}
   \begin{turn}{90}
   \begin{minipage}{1\textheight}
		\begin{center}
			\begin{tabular}{ | l | r ||  r | r | r | r | r | r | r || r | r | r | }
\hline
 & $\lambda$ [\AA]  & RED 1 & RED 2 & RED 3 & RED 4 & RED 5 & SF 1 & SF 2 & SF 3 & SF 4 & SF 5 \\
 \hline
 \hline
\ion{O}{ii} & 3727.09 & 2.924 & 1.803 & 1.621 & 1.992 & 2.323 & 28.630 & 21.238 & 13.033 & 9.077 & 7.940 \\
\ion{O}{ii} & 3729.88 & 1.577 & 1.254 & 1.200 & 1.445 & 1.531 & - & 26.363 & 16.573 & 11.115 & 9.387 \\
H$\theta$ & 3798.98 & - & - & - & - & - & - & - & - & - & - \\
H$\eta$ & 3836.47 & - & - & - & - & - & - & - & - & - & - \\
\ion{He}{i} & 3889.00 & - & - & - & - & - & 1.618 & 1.292 & 1.378 & 1.343 & 1.392 \\
\ion{S}{ii} & 4072.30 & - & - & - & - & - & - & - & - & - & - \\
H$\delta$ & 4102.89 & - & - & - & - & - & 2.706 & 2.103 & 2.069 & 1.920 & 1.842 \\
H$\gamma$ & 4341.68 & - & - & - & - & - & 6.111 & 4.408 & 3.916 & 3.457 & 3.166 \\
\ion{O}{iii} & 4364.44 & - & - & - & - & - & - & - & - & - & - \\
H$\beta$ & 4862.68 & 1.273 & - & - & - & - & 17.348 & 11.736 & 9.573 & 8.072 & 7.133 \\
\ion{O}{iii} & 4932.60 & - & - & - & - & - & - & - & - & - & - \\
\ion{O}{iii} & 4960.30 & - & - & - & - & - & 10.865 & 3.198 & 1.175 & - & - \\
\ion{O}{iii} & 5008.24 & - & - & - & - & - & 35.024 & 10.803 & 4.252 & 2.568 & 1.997 \\
\ion{O}{i} & 6302.05 & - & - & - & - & - & 2.400 & 1.600 & 1.239 & 1.107 & 1.072 \\
\ion{O}{i} & 6365.54 & - & - & - & - & - & - & - & - & - & - \\
\ion{N}{i} & 6529.03 & - & - & - & - & - & - & - & - & - & - \\
\ion{N}{ii} & 6549.86 & 2.276 & - & - & - & 1.038 & 4.771 & 4.464 & 4.372 & 4.355 & 5.050 \\
H$\alpha$ & 6564.61 & 6.305 & 2.206 & 1.123 & 1.409 & 1.958 & 84.884 & 53.510 & 41.972 & 36.237 & 33.311 \\
\ion{N}{ii} & 6585.27 & 3.229 & 1.598 & - & 1.281 & 1.657 & 13.877 & 13.506 & 13.443 & 13.223 & 13.223 \\
\ion{S}{ii} & 6718.29 & 1.276 & - & - & - & - & 12.821 & 9.729 & 7.407 & 5.949 & 5.322 \\
\ion{S}{ii} & 6732.67 & 1.033 & - & - & - & - & 9.335 & 6.959 & 5.312 & 4.353 & 4.001 \\
\hline
\end{tabular}

		\end{center} 
		 \caption{Values of the emission line equivalent widths (in \AA ngstr\"oms) for the refined galaxy classes.}
 		\label{tab:lines3}
 	\end{minipage}
 	\end{turn} 	
\end{table*}

\subsection{Continuum indices}  

We calculate numerous absorption line indices of the composites: Lick indices \citep{Worthey1994, Worthey1997}, BH indices (Brodie \& Hanes, see \citet{Huchra1996} and DTT indices \citep{Diaz1989}. Indices characterizing the size of the $ 4000 $ \AA \ break are from \citet{Bruzual1983} and \citet{Balogh1999}. Absorption indices for all composites are summarized in Tab.~\ref{tab:lick1}-\ref{tab:lick3} and are available online.

\begin{figure*}
   \begin{minipage}{1\textwidth}
    \begin{center}
	  \vspace{24pt}
	  \hspace{18pt}
      \input{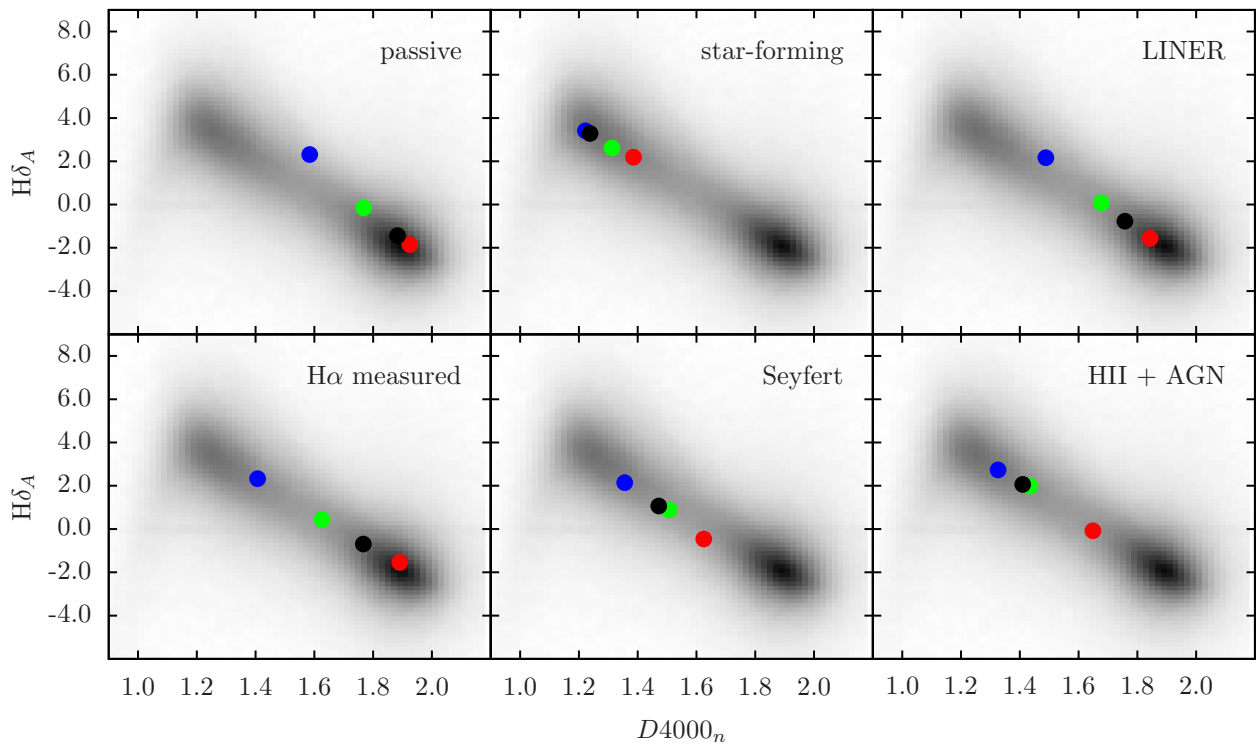}
      \vspace{36pt}
\caption{Recent star-formation signatures in the composites. The density plot is the square root of the number density of all SDSS galaxies. The loci of the composites are indicated with circles. The different colours denote the different colour classes; black points mark the loci of composites of all galaxies (regardless of their colours) in a given activity class. See the online edition for a colour version of this plot.}
      \label{fig:starburst}
    \end{center}
   \end{minipage}
  \end{figure*}

\begin{figure*}
   \begin{minipage}{1\textwidth}
    \begin{center}
	  \vspace{24pt}
	  \hspace{18pt}
      \input{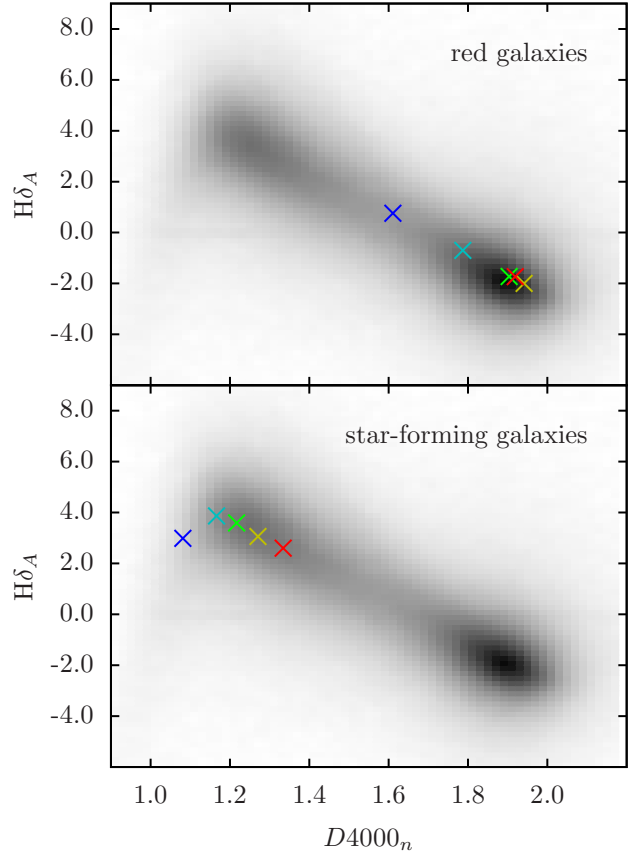}
      \vspace{36pt}
\caption{Recent star-formation signatures in the refined composites. The density plot is the square root of the number density of all SDSS galaxies. The crosses mark the loci of the composites based on the refined colour classification; see Sec.~\ref{sec:classrefine}.}
      \label{fig:starburst_refined}
    \end{center}
   \end{minipage}
  \end{figure*}

\textcolor{black}{One widely-used diagram to separate star-forming, starburst and post-starburst galaxies from passive ones based on absorption line features is the plot of the $D4000_n$ --H$\delta_A$ plane \citep{Kauffmann2003b, Kauffmann2003c}. As an illustration to the parameter space coverage of our composite atlas, we plot the $D4000_n$ -- H$\delta_A$ diagram in Fig.~\ref{fig:starburst}~and~\ref{fig:starburst_refined}. According to \citet{Kauffmann2003c}, in this plane, galaxies are located along a straight line: low-mass star-forming galaxies are close to the top left corner of the plot while high-mass passive galaxies are in the bottom right corner. Intermediate-mass galaxies ($10^{10} \leq M_\odot \leq 10^{11.5}$) show some bimodality and populate the two big clumps as well as the line between them. Starburst galaxies are located slightly above the straight line \citep{Kauffmann2003b}. The composites clearly follow the relationship among colour, H$ \alpha $ and H$ \delta_A $: passive red galaxies with no or weak H$ \alpha $ emissions are in the lower right corner, blue star-forming galaxies with strong H$ \alpha $ emissions are in the top left corner of the plots. Most AGNs fall into the intermediate-mass region, only the reddest LINERs overlapping with the clump of the completely passive red galaxies. In terms of recent star-formation activity, the two interesting galaxy classes are passive blue galaxies and blue LINERs. These galaxy classes are likely to contain many post-starburst galaxies as they are slightly above the linear distribution of the majority of galaxies.}

\begin{table*}
   \begin{turn}{90}
   \begin{minipage}{1\textheight}
		\begin{center}
			\begin{tabular}{ | l | c || r | r | r | r | r | r | r || r | r | r | r | r | r | r | }
\hline
 & & red & red & red & red & red & red & red & green & green & green & green & green & green & green \\
 & & P & H$\alpha$ & SF & A+\ion{H}{ii} & L & S & all & P & H$\alpha$ & SF & A+\ion{H}{ii} & L & S & all \\
 \hline
 \hline
CN$_1$ & mag & -0.033 & -0.038 & -0.062 & -0.044 & -0.055 & -0.031 & -0.041 & 0.050 & 0.032 & -0.060 & -0.029 & 0.045 & -0.004 & 0.033 \\
CN$_2$ & mag & 0.023 & 0.002 & -0.037 & -0.009 & -0.005 & 0.005 & -0.004 & 0.098 & 0.077 & -0.035 & 0.006 & 0.086 & 0.032 & 0.077 \\
Ca4227 & \AA & 0.875 & 0.785 & 0.447 & 0.615 & 0.728 & 0.663 & 0.720 & 1.260 & 1.172 & 0.468 & 0.744 & 1.160 & 0.861 & 1.147 \\
G4300 & \AA & 3.810 & 2.905 & 1.185 & 2.205 & 2.718 & 2.841 & 2.613 & 5.492 & 4.917 & 1.296 & 2.811 & 4.942 & 3.732 & 4.834 \\
Fe4383 & \AA & 3.467 & 3.124 & 1.640 & 2.698 & 3.120 & 3.595 & 2.928 & 4.945 & 4.638 & 1.737 & 3.173 & 4.814 & 4.351 & 4.576 \\
Ca4455 & \AA & 1.255 & 1.122 & 0.580 & 0.900 & 1.019 & 1.017 & 1.035 & 1.599 & 1.509 & 0.614 & 1.035 & 1.415 & 1.219 & 1.477 \\
Fe4531 & \AA & 3.156 & 2.720 & 1.913 & 2.384 & 2.614 & 2.644 & 2.591 & 3.511 & 3.336 & 1.965 & 2.612 & 3.309 & 2.983 & 3.302 \\
C4668 & \AA & 4.718 & 3.792 & 1.839 & 3.065 & 3.908 & 3.153 & 3.491 & 6.368 & 5.928 & 1.951 & 3.757 & 6.116 & 4.145 & 5.897 \\
H$\beta$ & \AA & 3.187 & 0.212 & -4.591 & -1.915 & 1.492 & -1.474 & -0.550 & 1.913 & 1.364 & -4.348 & -1.222 & 0.673 & -0.739 & 1.240 \\
Fe5015 & \AA & 4.695 & 3.583 & -0.179 & 0.914 & 1.673 & -7.702 & 2.748 & 5.008 & 4.525 & 0.155 & 1.825 & 2.987 & -5.896 & 4.369 \\
Mg$_1$ & mag & 0.055 & 0.051 & 0.025 & 0.048 & 0.059 & 0.067 & 0.049 & 0.101 & 0.093 & 0.027 & 0.058 & 0.104 & 0.079 & 0.094 \\
Mg$_2$ & mag & 0.159 & 0.145 & 0.086 & 0.127 & 0.149 & 0.150 & 0.139 & 0.238 & 0.223 & 0.090 & 0.150 & 0.234 & 0.178 & 0.224 \\
Mg $b$ & \AA & 2.671 & 2.646 & 1.812 & 2.557 & 2.980 & 2.923 & 2.585 & 3.839 & 3.686 & 1.884 & 2.887 & 4.113 & 3.347 & 3.684 \\
Fe5270 & \AA & 2.314 & 2.221 & 1.517 & 1.893 & 2.014 & 2.082 & 2.104 & 2.680 & 2.624 & 1.563 & 2.071 & 2.532 & 2.358 & 2.588 \\
Fe5335 & \AA & 2.159 & 2.144 & 1.543 & 1.907 & 2.019 & 2.057 & 2.050 & 2.382 & 2.370 & 1.581 & 2.017 & 2.310 & 2.260 & 2.327 \\
Fe5406 & \AA & 1.369 & 1.341 & 0.945 & 1.176 & 1.253 & 1.272 & 1.278 & 1.573 & 1.552 & 0.972 & 1.262 & 1.518 & 1.449 & 1.524 \\
Fe5709 & \AA & 0.823 & 0.784 & 0.594 & 0.701 & 0.760 & 0.803 & 0.754 & 0.906 & 0.890 & 0.611 & 0.758 & 0.871 & 0.874 & 0.877 \\
Fe5782 & \AA & 0.641 & 0.707 & 0.538 & 0.687 & 0.710 & 0.716 & 0.693 & 0.756 & 0.772 & 0.564 & 0.728 & 0.784 & 0.769 & 0.763 \\
Na D & \AA & 2.566 & 2.209 & 1.282 & 2.363 & 2.447 & 2.121 & 2.208 & 3.366 & 3.266 & 1.495 & 2.864 & 3.471 & 2.678 & 3.302 \\
TiO$_1$ & mag & 0.022 & 0.022 & 0.016 & 0.020 & 0.024 & 0.025 & 0.022 & 0.032 & 0.031 & 0.016 & 0.023 & 0.032 & 0.027 & 0.031 \\
TiO$_2$ & mag & 0.053 & 0.053 & 0.041 & 0.051 & 0.056 & 0.056 & 0.052 & 0.073 & 0.070 & 0.042 & 0.056 & 0.074 & 0.062 & 0.070 \\
H$\delta_A$ & \AA & 2.656 & 2.405 & 3.501 & 2.829 & 3.544 & 2.447 & 2.608 & -1.280 & -0.514 & 3.409 & 2.176 & -0.668 & 1.180 & -0.445 \\
H$\gamma_A$ & \AA & -0.344 & -0.822 & 0.128 & -0.668 & 0.334 & -1.528 & -0.705 & -4.969 & -4.202 & 0.026 & -1.476 & -4.852 & -2.984 & -4.172 \\
H$\delta_F$ & \AA & 2.676 & 2.265 & 2.411 & 2.261 & 3.096 & 2.084 & 2.287 & 0.625 & 0.934 & 2.373 & 1.973 & 0.823 & 1.501 & 0.944 \\
H$\gamma_F$ & \AA & 1.562 & 0.643 & 0.225 & 0.347 & 1.653 & 0.101 & 0.559 & -1.092 & -0.865 & 0.211 & 0.035 & -1.279 & -0.531 & -0.867 \\
\ion{Ca}{ii} $\lambda8498$ & \AA & 0.794 & 0.854 & 0.757 & 0.820 & 0.806 & 0.842 & 0.842 & 1.150 & 1.155 & 0.775 & 1.001 & 1.122 & 0.880 & 1.147 \\
\ion{Ca}{ii} $\lambda8542$ & \AA & 2.723 & 2.741 & 2.488 & 2.673 & 2.615 & 2.710 & 2.720 & 3.334 & 3.342 & 2.514 & 2.939 & 3.259 & 2.757 & 3.323 \\
\ion{Ca}{ii} $\lambda8662$ & \AA & 2.719 & 2.802 & 2.683 & 2.777 & 2.696 & 2.779 & 2.798 & 3.061 & 3.082 & 2.691 & 2.940 & 2.982 & 2.759 & 3.062 \\
\ion{Mg}{i} $\lambda8807$ & \AA & 0.448 & 0.473 & 0.343 & 0.415 & 0.382 & 0.449 & 0.460 & 0.372 & 0.382 & 0.359 & 0.343 & 0.339 & 0.465 & 0.371 \\
CNB & mag & 0.115 & 0.080 & 0.028 & 0.060 & 0.087 & 0.051 & 0.071 & 0.202 & 0.166 & 0.030 & 0.073 & 0.161 & 0.080 & 0.163 \\
H + K & mag & 0.333 & 0.248 & 0.183 & 0.216 & 0.258 & 0.229 & 0.235 & 0.373 & 0.336 & 0.187 & 0.239 & 0.329 & 0.263 & 0.331 \\
\ion{Ca}{i} & mag & 0.007 & 0.009 & 0.006 & 0.007 & 0.006 & 0.005 & 0.008 & 0.015 & 0.014 & 0.007 & 0.009 & 0.013 & 0.007 & 0.014 \\
G & mag & 0.173 & 0.141 & 0.073 & 0.110 & 0.134 & 0.134 & 0.128 & 0.232 & 0.211 & 0.077 & 0.132 & 0.207 & 0.168 & 0.206 \\
Hb & mag & 0.104 & 0.046 & -0.036 & 0.007 & 0.073 & 0.013 & 0.032 & 0.073 & 0.064 & -0.032 & 0.020 & 0.051 & 0.026 & 0.062 \\
MgG & mag & 0.067 & 0.064 & 0.040 & 0.060 & 0.071 & 0.067 & 0.062 & 0.102 & 0.096 & 0.042 & 0.069 & 0.107 & 0.079 & 0.096 \\
MH & mag & 0.034 & 0.036 & 0.018 & 0.024 & 0.023 & -0.028 & 0.032 & 0.060 & 0.055 & 0.020 & 0.031 & 0.051 & -0.014 & 0.055 \\
FC & mag & 0.062 & 0.056 & 0.037 & 0.049 & 0.051 & 0.052 & 0.053 & 0.073 & 0.070 & 0.038 & 0.054 & 0.069 & 0.060 & 0.070 \\
NaD & mag & 0.057 & 0.044 & 0.014 & 0.044 & 0.053 & 0.039 & 0.043 & 0.080 & 0.075 & 0.019 & 0.057 & 0.080 & 0.053 & 0.076 \\
D$4000$ & - & 1.785 & 1.572 & 1.348 & 1.481 & 1.617 & 1.543 & 1.533 & 2.070 & 1.938 & 1.364 & 1.576 & 1.971 & 1.677 & 1.924 \\
D$4000_n$ & - & 1.578 & 1.401 & 1.223 & 1.320 & 1.427 & 1.365 & 1.366 & 1.868 & 1.737 & 1.234 & 1.397 & 1.758 & 1.482 & 1.724 \\
\hline
\end{tabular}

		\end{center} 
		 \caption{Values of the continuum absorption indices for every galaxy class, continued in Tab.~\ref{tab:lick2}.}
 		\label{tab:lick1}
 	\end{minipage}
 	\end{turn} 	
\end{table*}  

\begin{table*}
   \begin{turn}{90}
   \begin{minipage}{1\textheight}
		\begin{center}
			\begin{tabular}{ | l | c || r | r | r | r | r | r | r || r | r | r | r | r | r | r | }
\hline
 & & blue & blue & blue & blue & blue & blue & blue & all & all & all & all & all & all & all \\
 & & P & H$\alpha$ & SF & A+\ion{H}{ii} & L & S & all & P & H$\alpha$ & SF & A+\ion{H}{ii} & L & S & all \\
 \hline
 \hline
CN$_1$ & mag & -0.033 & -0.038 & -0.062 & -0.044 & -0.055 & -0.031 & -0.041 & 0.050 & 0.032 & -0.060 & -0.029 & 0.045 & -0.004 & 0.033 \\
CN$_2$ & mag & 0.023 & 0.002 & -0.037 & -0.009 & -0.005 & 0.005 & -0.004 & 0.098 & 0.077 & -0.035 & 0.006 & 0.086 & 0.032 & 0.077 \\
Ca4227 & \AA & 0.875 & 0.785 & 0.447 & 0.615 & 0.728 & 0.663 & 0.720 & 1.260 & 1.172 & 0.468 & 0.744 & 1.160 & 0.861 & 1.147 \\
G4300 & \AA & 3.810 & 2.905 & 1.185 & 2.205 & 2.718 & 2.841 & 2.613 & 5.492 & 4.917 & 1.296 & 2.811 & 4.942 & 3.732 & 4.834 \\
Fe4383 & \AA & 3.467 & 3.124 & 1.640 & 2.698 & 3.120 & 3.595 & 2.928 & 4.945 & 4.638 & 1.737 & 3.173 & 4.814 & 4.351 & 4.576 \\
Ca4455 & \AA & 1.255 & 1.122 & 0.580 & 0.900 & 1.019 & 1.017 & 1.035 & 1.599 & 1.509 & 0.614 & 1.035 & 1.415 & 1.219 & 1.477 \\
Fe4531 & \AA & 3.156 & 2.720 & 1.913 & 2.384 & 2.614 & 2.644 & 2.591 & 3.511 & 3.336 & 1.965 & 2.612 & 3.309 & 2.983 & 3.302 \\
C4668 & \AA & 4.718 & 3.792 & 1.839 & 3.065 & 3.908 & 3.153 & 3.491 & 6.368 & 5.928 & 1.951 & 3.757 & 6.116 & 4.145 & 5.897 \\
H$\beta$ & \AA & 3.187 & 0.212 & -4.591 & -1.915 & 1.492 & -1.474 & -0.550 & 1.913 & 1.364 & -4.348 & -1.222 & 0.673 & -0.739 & 1.240 \\
Fe5015 & \AA & 4.695 & 3.583 & -0.179 & 0.914 & 1.673 & -7.702 & 2.748 & 5.008 & 4.525 & 0.155 & 1.825 & 2.987 & -5.896 & 4.369 \\
Mg$_1$ & mag & 0.055 & 0.051 & 0.025 & 0.048 & 0.059 & 0.067 & 0.049 & 0.101 & 0.093 & 0.027 & 0.058 & 0.104 & 0.079 & 0.094 \\
Mg$_2$ & mag & 0.159 & 0.145 & 0.086 & 0.127 & 0.149 & 0.150 & 0.139 & 0.238 & 0.223 & 0.090 & 0.150 & 0.234 & 0.178 & 0.224 \\
Mg $b$ & \AA & 2.671 & 2.646 & 1.812 & 2.557 & 2.980 & 2.923 & 2.585 & 3.839 & 3.686 & 1.884 & 2.887 & 4.113 & 3.347 & 3.684 \\
Fe5270 & \AA & 2.314 & 2.221 & 1.517 & 1.893 & 2.014 & 2.082 & 2.104 & 2.680 & 2.624 & 1.563 & 2.071 & 2.532 & 2.358 & 2.588 \\
Fe5335 & \AA & 2.159 & 2.144 & 1.543 & 1.907 & 2.019 & 2.057 & 2.050 & 2.382 & 2.370 & 1.581 & 2.017 & 2.310 & 2.260 & 2.327 \\
Fe5406 & \AA & 1.369 & 1.341 & 0.945 & 1.176 & 1.253 & 1.272 & 1.278 & 1.573 & 1.552 & 0.972 & 1.262 & 1.518 & 1.449 & 1.524 \\
Fe5709 & \AA & 0.823 & 0.784 & 0.594 & 0.701 & 0.760 & 0.803 & 0.754 & 0.906 & 0.890 & 0.611 & 0.758 & 0.871 & 0.874 & 0.877 \\
Fe5782 & \AA & 0.641 & 0.707 & 0.538 & 0.687 & 0.710 & 0.716 & 0.693 & 0.756 & 0.772 & 0.564 & 0.728 & 0.784 & 0.769 & 0.763 \\
Na D & \AA & 2.566 & 2.209 & 1.282 & 2.363 & 2.447 & 2.121 & 2.208 & 3.366 & 3.266 & 1.495 & 2.864 & 3.471 & 2.678 & 3.302 \\
TiO$_1$ & mag & 0.022 & 0.022 & 0.016 & 0.020 & 0.024 & 0.025 & 0.022 & 0.032 & 0.031 & 0.016 & 0.023 & 0.032 & 0.027 & 0.031 \\
TiO$_2$ & mag & 0.053 & 0.053 & 0.041 & 0.051 & 0.056 & 0.056 & 0.052 & 0.073 & 0.070 & 0.042 & 0.056 & 0.074 & 0.062 & 0.070 \\
H$\delta_A$ & \AA & 2.656 & 2.405 & 3.501 & 2.829 & 3.544 & 2.447 & 2.608 & -1.280 & -0.514 & 3.409 & 2.176 & -0.668 & 1.180 & -0.445 \\
H$\gamma_A$ & \AA & -0.344 & -0.822 & 0.128 & -0.668 & 0.334 & -1.528 & -0.705 & -4.969 & -4.202 & 0.026 & -1.476 & -4.852 & -2.984 & -4.172 \\
H$\delta_F$ & \AA & 2.676 & 2.265 & 2.411 & 2.261 & 3.096 & 2.084 & 2.287 & 0.625 & 0.934 & 2.373 & 1.973 & 0.823 & 1.501 & 0.944 \\
H$\gamma_F$ & \AA & 1.562 & 0.643 & 0.225 & 0.347 & 1.653 & 0.101 & 0.559 & -1.092 & -0.865 & 0.211 & 0.035 & -1.279 & -0.531 & -0.867 \\
\ion{Ca}{ii} $\lambda8498$ & \AA & 0.794 & 0.854 & 0.757 & 0.820 & 0.806 & 0.842 & 0.842 & 1.150 & 1.155 & 0.775 & 1.001 & 1.122 & 0.880 & 1.147 \\
\ion{Ca}{ii} $\lambda8542$ & \AA & 2.723 & 2.741 & 2.488 & 2.673 & 2.615 & 2.710 & 2.720 & 3.334 & 3.342 & 2.514 & 2.939 & 3.259 & 2.757 & 3.323 \\
\ion{Ca}{ii} $\lambda8662$ & \AA & 2.719 & 2.802 & 2.683 & 2.777 & 2.696 & 2.779 & 2.798 & 3.061 & 3.082 & 2.691 & 2.940 & 2.982 & 2.759 & 3.062 \\
\ion{Mg}{i} $\lambda8807$ & \AA & 0.448 & 0.473 & 0.343 & 0.415 & 0.382 & 0.449 & 0.460 & 0.372 & 0.382 & 0.359 & 0.343 & 0.339 & 0.465 & 0.371 \\
CNB & mag & 0.115 & 0.080 & 0.028 & 0.060 & 0.087 & 0.051 & 0.071 & 0.202 & 0.166 & 0.030 & 0.073 & 0.161 & 0.080 & 0.163 \\
H + K & mag & 0.333 & 0.248 & 0.183 & 0.216 & 0.258 & 0.229 & 0.235 & 0.373 & 0.336 & 0.187 & 0.239 & 0.329 & 0.263 & 0.331 \\
\ion{Ca}{i} & mag & 0.007 & 0.009 & 0.006 & 0.007 & 0.006 & 0.005 & 0.008 & 0.015 & 0.014 & 0.007 & 0.009 & 0.013 & 0.007 & 0.014 \\
G & mag & 0.173 & 0.141 & 0.073 & 0.110 & 0.134 & 0.134 & 0.128 & 0.232 & 0.211 & 0.077 & 0.132 & 0.207 & 0.168 & 0.206 \\
Hb & mag & 0.104 & 0.046 & -0.036 & 0.007 & 0.073 & 0.013 & 0.032 & 0.073 & 0.064 & -0.032 & 0.020 & 0.051 & 0.026 & 0.062 \\
MgG & mag & 0.067 & 0.064 & 0.040 & 0.060 & 0.071 & 0.067 & 0.062 & 0.102 & 0.096 & 0.042 & 0.069 & 0.107 & 0.079 & 0.096 \\
MH & mag & 0.034 & 0.036 & 0.018 & 0.024 & 0.023 & -0.028 & 0.032 & 0.060 & 0.055 & 0.020 & 0.031 & 0.051 & -0.014 & 0.055 \\
FC & mag & 0.062 & 0.056 & 0.037 & 0.049 & 0.051 & 0.052 & 0.053 & 0.073 & 0.070 & 0.038 & 0.054 & 0.069 & 0.060 & 0.070 \\
NaD & mag & 0.057 & 0.044 & 0.014 & 0.044 & 0.053 & 0.039 & 0.043 & 0.080 & 0.075 & 0.019 & 0.057 & 0.080 & 0.053 & 0.076 \\
D$4000$ & - & 1.785 & 1.572 & 1.348 & 1.481 & 1.617 & 1.543 & 1.533 & 2.070 & 1.938 & 1.364 & 1.576 & 1.971 & 1.677 & 1.924 \\
D$4000_n$ & - & 1.578 & 1.401 & 1.223 & 1.320 & 1.427 & 1.365 & 1.366 & 1.868 & 1.737 & 1.234 & 1.397 & 1.758 & 1.482 & 1.724 \\
\hline
\end{tabular}

		\end{center} 
		 \caption{Values of the continuum indices for every galaxy class, continued from Tab.~\ref{tab:lick1}.}
 		\label{tab:lick2}
 	\end{minipage}
 	\end{turn} 	
\end{table*} 

\begin{table*}
   \begin{turn}{90}
   \begin{minipage}{1\textheight}
		\begin{center}
			\begin{tabular}{ | l | c || r | r | r | r | r | r | r || r | r | r | }
\hline
 & & RED 1 & RED 2 & RED 3 & RED 4 & RED 5 & SF 1 & SF 2 & SF 3 & SF 4 & SF 5 \\
 \hline
 \hline
CN$_1$ & mag & -0.002 & 0.032 & 0.063 & 0.079 & 0.066 & -0.047 & -0.069 & -0.064 & -0.054 & -0.047 \\
CN$_2$ & mag & 0.043 & 0.078 & 0.110 & 0.124 & 0.110 & -0.045 & -0.049 & -0.039 & -0.027 & -0.019 \\
Ca4227 & \AA & 1.038 & 1.209 & 1.304 & 1.301 & 1.286 & 0.205 & 0.341 & 0.429 & 0.539 & 0.622 \\
G4300 & \AA & 4.289 & 5.175 & 5.563 & 5.580 & 5.465 & 0.092 & 0.597 & 1.130 & 1.659 & 2.095 \\
Fe4383 & \AA & 4.177 & 4.809 & 5.087 & 5.108 & 5.046 & 1.123 & 1.194 & 1.576 & 2.064 & 2.461 \\
Ca4455 & \AA & 1.418 & 1.575 & 1.604 & 1.574 & 1.536 & 0.096 & 0.361 & 0.576 & 0.735 & 0.854 \\
Fe4531 & \AA & 3.190 & 3.426 & 3.499 & 3.483 & 3.430 & 1.183 & 1.625 & 1.890 & 2.122 & 2.318 \\
C4668 & \AA & 5.160 & 6.034 & 6.615 & 6.927 & 6.669 & 0.683 & 1.198 & 1.751 & 2.337 & 2.773 \\
H$\beta$ & \AA & 1.407 & 1.734 & 1.597 & 1.434 & 1.409 & -12.371 & -6.644 & -4.584 & -3.640 & -3.117 \\
Fe5015 & \AA & 4.305 & 4.728 & 4.825 & 4.741 & 4.567 & -16.711 & -3.991 & 0.305 & 1.743 & 2.286 \\
Mg$_1$ & mag & 0.072 & 0.089 & 0.109 & 0.120 & 0.113 & 0.019 & 0.019 & 0.024 & 0.032 & 0.039 \\
Mg$_2$ & mag & 0.186 & 0.218 & 0.251 & 0.268 & 0.256 & 0.054 & 0.068 & 0.083 & 0.101 & 0.115 \\
Mg $b$ & \AA & 3.210 & 3.626 & 4.039 & 4.232 & 4.085 & 1.301 & 1.520 & 1.763 & 2.080 & 2.333 \\
Fe5270 & \AA & 2.544 & 2.690 & 2.706 & 2.690 & 2.669 & 0.800 & 1.229 & 1.495 & 1.710 & 1.842 \\
Fe5335 & \AA & 2.370 & 2.447 & 2.380 & 2.316 & 2.320 & 0.963 & 1.295 & 1.525 & 1.708 & 1.802 \\
Fe5406 & \AA & 1.525 & 1.594 & 1.582 & 1.543 & 1.525 & 0.594 & 0.800 & 0.933 & 1.054 & 1.130 \\
Fe5709 & \AA & 0.889 & 0.923 & 0.901 & 0.878 & 0.897 & 0.376 & 0.505 & 0.578 & 0.654 & 0.705 \\
Fe5782 & \AA & 0.770 & 0.780 & 0.774 & 0.776 & 0.785 & 0.303 & 0.412 & 0.523 & 0.631 & 0.716 \\
Na D & \AA & 2.732 & 3.077 & 3.575 & 4.066 & 4.481 & 0.015 & 0.633 & 1.162 & 1.898 & 2.826 \\
TiO$_1$ & mag & 0.026 & 0.030 & 0.033 & 0.035 & 0.034 & 0.012 & 0.014 & 0.016 & 0.017 & 0.018 \\
TiO$_2$ & mag & 0.062 & 0.069 & 0.076 & 0.080 & 0.078 & 0.032 & 0.034 & 0.040 & 0.046 & 0.050 \\
H$\delta_A$ & \AA & 0.762 & -0.698 & -1.724 & -1.993 & -1.728 & 3.012 & 3.859 & 3.591 & 3.057 & 2.602 \\
H$\gamma_A$ & \AA & -2.713 & -4.343 & -5.508 & -5.856 & -5.574 & -1.284 & 0.332 & 0.260 & -0.311 & -0.868 \\
H$\delta_F$ & \AA & 1.568 & 0.893 & 0.379 & 0.222 & 0.341 & 1.653 & 2.489 & 2.463 & 2.217 & 1.986 \\
H$\gamma_F$ & \AA & -0.081 & -0.834 & -1.457 & -1.655 & -1.514 & -1.718 & -0.004 & 0.301 & 0.169 & -0.037 \\
\ion{Ca}{ii} $\lambda8498$ & \AA & 1.078 & 1.167 & 1.144 & 1.127 & 1.187 & 0.470 & 0.561 & 0.752 & 0.807 & 0.814 \\
\ion{Ca}{ii} $\lambda8542$ & \AA & 3.231 & 3.370 & 3.318 & 3.272 & 3.445 & 1.967 & 2.166 & 2.476 & 2.581 & 2.556 \\
\ion{Ca}{ii} $\lambda8662$ & \AA & 3.074 & 3.114 & 3.010 & 2.961 & 3.182 & 2.329 & 2.476 & 2.682 & 2.728 & 2.676 \\
\ion{Mg}{i} $\lambda8807$ & \AA & 0.410 & 0.392 & 0.365 & 0.342 & 0.478 & 0.101 & 0.200 & 0.342 & 0.408 & 0.421 \\
CNB & mag & 0.125 & 0.175 & 0.213 & 0.229 & 0.214 & -0.006 & 0.021 & 0.027 & 0.037 & 0.046 \\
H + K & mag & 0.310 & 0.352 & 0.372 & 0.375 & 0.368 & 0.113 & 0.163 & 0.182 & 0.198 & 0.218 \\
\ion{Ca}{i} & mag & 0.012 & 0.013 & 0.016 & 0.016 & 0.016 & 0.002 & 0.005 & 0.006 & 0.007 & 0.008 \\
G & mag & 0.193 & 0.224 & 0.232 & 0.227 & 0.224 & 0.030 & 0.052 & 0.071 & 0.090 & 0.106 \\
Hb & mag & 0.067 & 0.071 & 0.067 & 0.065 & 0.065 & -0.166 & -0.070 & -0.036 & -0.020 & -0.010 \\
MgG & mag & 0.080 & 0.093 & 0.108 & 0.116 & 0.110 & 0.026 & 0.033 & 0.039 & 0.047 & 0.054 \\
MH & mag & 0.044 & 0.054 & 0.063 & 0.068 & 0.065 & -0.064 & 0.000 & 0.020 & 0.028 & 0.033 \\
FC & mag & 0.066 & 0.071 & 0.074 & 0.075 & 0.074 & 0.017 & 0.029 & 0.037 & 0.043 & 0.046 \\
NaD & mag & 0.060 & 0.071 & 0.085 & 0.098 & 0.105 & -0.031 & -0.006 & 0.011 & 0.029 & 0.050 \\
D$4000$ & - & 1.814 & 1.993 & 2.105 & 2.146 & 2.145 & 1.161 & 1.273 & 1.342 & 1.415 & 1.514 \\
D$4000_n$ & - & 1.611 & 1.787 & 1.903 & 1.941 & 1.919 & 1.081 & 1.166 & 1.217 & 1.271 & 1.334 \\
\hline
\end{tabular}

		\end{center} 
		 \caption{Value of the continuum absorption indices for the refined colour classes.}
		\label{tab:lick3}
 	\end{minipage}
 	\end{turn} 	
\end{table*} 

\subsection{Physical parameters from Bayesian analysis}
\label{params}

The model fitting method described in Sec.~\ref{sec:models} is known not to yield correct values for the physical parameters of the stellar populations building up the galaxies, and more sophisticated techniques are required to estimate the age and the metallicity of the galaxies.

\textcolor{black}{To estimate the physical parameters of the continua of the composites, we entirely adopt the method described in Sec.~4 of \citet{Kauffmann2003b}. They used a Bayesian approach to sample the parameter space of the BC03 stellar population synthesis models \citep{BC03} paying special attention to generate physically plausible star formation histories and constrain other parameters (metallicity, dust content). Based on the stochastic library of star formation histories, they generated a huge library of model spectra. The star formation histories were characterized by a continuous rate starting at $t_{\text{form}}$ and decreasing exponentially with the time-scale $\gamma$ superimposed by random, short-term bursts.}

\textcolor{black}{To estimate physical parameters of a single spectrum, models of the library are individually fitted to the composites and from the goodness of these fits a probability of the model parameter set is calculated. This method gives not just a single best-fitting model to the spectrum as the NNLS method described in Sec.~\ref{sec:models} does, but yields the whole probability distribution of the model parameters and physical properties. As certain physical parameters are very hard to constrain based on optical spectroscopy, it is much better to characterize galaxy spectra with the probability distribution of their parameters instead of a single best-fit model. The distributions of the physical parameters of our composites are determined using the described method and the modes and medians of the distributions are listed in Tab.~\ref{tab:contpar}. It is clear from the results that our composites cover a very broad range of stellar population ages, specific star formation rates and mass-to-light ratios. On the other hand, metallicities are weakly constrained and close to solar. For the same parameter measurements of individual SDSS galaxies, see \citet{Brinchmann2004}.}

\begin{table*}
   \begin{turn}{90}
   \begin{minipage}{1\textheight}
		\begin{center}
			\begin{tabular}{ | l || r | r | r | r | r | r | r | r | r | r | r | r | r | r | r | r | }
\hline
 & \multicolumn{2}{|c|}{$t_{\text{form}}$ [Gyr]} & \multicolumn{2}{|c|}{$\gamma$ [Gyr]} & \multicolumn{2}{|c|}{$t_M$ [Gyr]} & \multicolumn{2}{|c|}{$t_r$ [Gyr]} & \multicolumn{2}{|c|}{$Z$ [$Z_\odot$]} & \multicolumn{2}{|c|}{$\tau_V$} & \multicolumn{2}{|c|}{$\log{\text{SSFR}}$} & \multicolumn{2}{|c|}{$\log{ \frac{M}{L} }$} \\
 & median & mode & median & mode & median & mode & median & mode & median & mode & median & mode & median & mode & median & mode \\
\hline
\hline
P & 10.33 & 11.42 & 0.68 & 0.89 & 6.42 & 6.42 & 5.31 & 4.57 & 1.09 & 0.89 & 0.80 & 0.06 & -12.11 & -11.40 & 0.49 & 0.49 \\
H$\alpha$ & 10.33 & 11.69 & 0.66 & 0.97 & 6.42 & 6.67 & 5.55 & 4.57 & 1.21 & 1.09 & 1.16 & 0.67 & -11.93 & -11.40 & 0.57 & 0.57 \\
SF & 8.69 & 10.88 & 0.50 & 0.81 & 5.17 & 4.17 & 4.08 & 3.10 & 1.09 & 1.71 & 1.90 & 1.53 & -11.05 & -10.70 & 0.57 & 0.57 \\
A+\ion{H}{ii} & 9.78 & 12.78 & 0.60 & 0.85 & 5.92 & 5.42 & 5.06 & 4.57 & 1.24 & 1.71 & 1.53 & 1.28 & -11.58 & -11.05 & 0.57 & 0.57 \\
L & 10.33 & 12.78 & 0.66 & 0.85 & 6.67 & 6.67 & 5.55 & 4.57 & 1.28 & 1.55 & 1.28 & 1.28 & -11.93 & -11.23 & 0.57 & 0.57 \\
S & 9.78 & 12.78 & 0.62 & 0.77 & 6.17 & 5.42 & 5.06 & 4.57 & 1.17 & 1.71 & 1.41 & 1.28 & -11.58 & -11.05 & 0.57 & 0.57 \\
all & 10.33 & 11.97 & 0.68 & 0.97 & 6.42 & 6.67 & 5.55 & 4.57 & 1.21 & 1.09 & 1.04 & 0.67 & -12.11 & -11.40 & 0.57 & 0.57 \\
P & 8.97 & 10.88 & 0.62 & 0.83 & 5.17 & 4.67 & 3.83 & 3.34 & 0.89 & 0.62 & 0.80 & 0.06 & -11.40 & -11.05 & 0.31 & 0.31 \\
H$\alpha$ & 9.24 & 11.97 & 0.58 & 0.77 & 5.42 & 4.67 & 4.32 & 3.83 & 1.01 & 0.55 & 1.28 & 1.16 & -11.23 & -10.88 & 0.40 & 0.40 \\
SF & 7.33 & 4.87 & 0.44 & 0.13 & 4.17 & 3.92 & 3.10 & 2.36 & 1.01 & 0.55 & 1.65 & 1.28 & -10.53 & -10.35 & 0.40 & 0.31 \\
A+\ion{H}{ii} & 8.15 & 6.24 & 0.50 & 0.30 & 4.92 & 4.67 & 3.83 & 3.34 & 0.97 & 0.55 & 1.53 & 1.16 & -10.88 & -10.53 & 0.40 & 0.40 \\
L & 9.51 & 11.97 & 0.58 & 0.77 & 5.67 & 4.67 & 4.57 & 3.83 & 1.01 & 0.74 & 1.28 & 1.16 & -11.40 & -11.05 & 0.49 & 0.49 \\
S & 8.69 & 6.24 & 0.54 & 0.77 & 5.17 & 4.67 & 4.08 & 3.83 & 0.97 & 0.55 & 1.41 & 1.16 & -11.05 & -10.70 & 0.40 & 0.40 \\
all & 9.24 & 11.97 & 0.58 & 0.77 & 5.42 & 4.67 & 4.32 & 3.83 & 1.01 & 0.55 & 1.16 & 1.16 & -11.40 & -10.88 & 0.40 & 0.40 \\
P & 7.87 & 6.78 & 0.54 & 0.91 & 4.17 & 3.42 & 2.85 & 2.12 & 0.62 & 0.20 & 0.92 & 0.06 & -11.05 & -10.53 & 0.14 & 0.14 \\
H$\alpha$ & 7.33 & 4.87 & 0.44 & 0.32 & 4.17 & 3.92 & 3.10 & 2.36 & 0.86 & 0.16 & 1.28 & 1.16 & -10.53 & -10.35 & 0.31 & 0.31 \\
SF & 4.33 & 3.24 & 0.34 & 0.01 & 2.42 & 1.92 & 1.63 & 1.38 & 0.93 & 0.12 & 1.28 & 1.16 & -9.82 & -9.82 & -0.03 & -0.03 \\
A+\ion{H}{ii} & 6.51 & 4.87 & 0.38 & 0.03 & 3.92 & 2.92 & 2.85 & 2.12 & 0.93 & 0.16 & 1.41 & 1.28 & -10.35 & -10.35 & 0.23 & 0.23 \\
L & 7.33 & 5.42 & 0.46 & 0.32 & 4.17 & 3.42 & 3.10 & 2.85 & 0.82 & 0.16 & 1.16 & 1.16 & -10.70 & -10.53 & 0.23 & 0.23 \\
S & 7.06 & 4.87 & 0.42 & 0.13 & 4.17 & 3.42 & 3.10 & 2.36 & 0.89 & 0.16 & 1.28 & 1.16 & -10.53 & -10.35 & 0.31 & 0.31 \\
all & 6.78 & 4.87 & 0.40 & 0.32 & 4.17 & 3.42 & 3.10 & 2.36 & 0.89 & 0.16 & 1.28 & 1.28 & -10.53 & -10.35 & 0.23 & 0.23 \\
P & 10.06 & 11.15 & 0.66 & 0.89 & 6.17 & 6.42 & 5.06 & 4.57 & 1.09 & 0.89 & 0.80 & 0.06 & -11.93 & -11.23 & 0.49 & 0.49 \\
H$\alpha$ & 9.78 & 11.15 & 0.62 & 0.77 & 5.92 & 4.67 & 4.82 & 3.83 & 1.13 & 1.09 & 1.16 & 1.16 & -11.58 & -11.05 & 0.49 & 0.49 \\
SF & 4.87 & 3.24 & 0.34 & 0.01 & 2.67 & 2.17 & 1.87 & 1.63 & 0.93 & 0.12 & 1.28 & 1.16 & -10.00 & -9.82 & 0.06 & 0.06 \\
A+\ion{H}{ii} & 7.87 & 5.69 & 0.48 & 0.30 & 4.67 & 3.92 & 3.59 & 2.85 & 0.97 & 0.55 & 1.41 & 1.16 & -10.70 & -10.53 & 0.40 & 0.40 \\
L & 9.78 & 12.78 & 0.62 & 0.85 & 6.17 & 6.67 & 5.06 & 4.57 & 1.24 & 1.52 & 1.41 & 1.28 & -11.76 & -11.05 & 0.57 & 0.57 \\
S & 8.69 & 6.24 & 0.54 & 0.30 & 5.17 & 4.67 & 4.08 & 3.83 & 1.01 & 0.55 & 1.41 & 1.16 & -11.05 & -10.70 & 0.40 & 0.40 \\
all & 9.78 & 11.15 & 0.62 & 0.77 & 5.92 & 4.67 & 4.82 & 3.83 & 1.13 & 1.09 & 1.16 & 1.16 & -11.58 & -11.05 & 0.49 & 0.49 \\
RED 1 & 9.24 & 10.88 & 0.58 & 0.77 & 5.42 & 4.67 & 4.32 & 3.83 & 0.97 & 0.55 & 1.16 & 1.16 & -11.23 & -10.88 & 0.40 & 0.40 \\
RED 2 & 10.06 & 11.15 & 0.64 & 0.77 & 5.92 & 4.67 & 4.82 & 3.83 & 1.09 & 0.89 & 1.04 & 0.67 & -11.76 & -11.23 & 0.49 & 0.49 \\
RED 3 & 10.33 & 11.97 & 0.68 & 0.97 & 6.42 & 6.67 & 5.55 & 4.57 & 1.17 & 1.09 & 1.04 & 0.67 & -12.11 & -11.40 & 0.49 & 0.49 \\
RED 4 & 10.60 & 12.78 & 0.70 & 0.83 & 6.92 & 6.67 & 6.04 & 5.55 & 1.28 & 1.17 & 1.16 & 1.28 & -12.11 & -11.76 & 0.57 & 0.57 \\
RED 5 & 10.60 & 12.78 & 0.68 & 0.85 & 7.17 & 7.92 & 6.29 & 5.55 & 1.48 & 1.94 & 1.53 & 1.53 & -12.11 & -11.76 & 0.74 & 0.74 \\
SF 1 & 1.33 & 1.06 & 0.44 & 0.56 & 0.67 & 0.42 & 0.40 & 0.40 & 0.86 & 0.78 & 1.16 & 0.92 & -8.94 & -8.94 & -0.63 & -0.63 \\
SF 2 & 2.69 & 1.87 & 0.36 & 0.01 & 1.42 & 0.92 & 0.89 & 0.89 & 0.93 & 0.12 & 1.16 & 0.80 & -9.47 & -9.47 & -0.28 & -0.37 \\
SF 3 & 4.06 & 2.96 & 0.34 & 0.01 & 2.42 & 1.92 & 1.63 & 1.38 & 0.97 & 0.12 & 1.28 & 1.16 & -9.82 & -9.82 & -0.03 & -0.03 \\
SF 4 & 5.69 & 3.51 & 0.36 & 0.13 & 3.42 & 2.67 & 2.36 & 1.63 & 1.01 & 0.78 & 1.41 & 1.16 & -10.17 & -10.00 & 0.23 & 0.23 \\
SF 5 & 7.60 & 4.87 & 0.46 & 0.30 & 4.67 & 3.92 & 3.34 & 2.36 & 1.01 & 0.55 & 1.65 & 1.53 & -10.70 & -10.35 & 0.40 & 0.40 \\
\hline
\end{tabular}

		\end{center} 
		 \caption{Modes and medians of the probability distributions of the physical parameters of the composites resulting from the Bayesian continuum fitting technique. $t_\text{form}$: age of the continuous model component; $ \gamma $: time-scale parameter of the star formation rate; $ t_M $: average age of the stellar population, weighted by the stellar mass; $ t_r $: average age of the stellar population, weighted by the r-band magnitude; $ Z $: metallicity; $ \tau_V $: optical depth of the dust component; $ \text{SSFR} $: specific star formation rate; $ M/L $: mass-to-light ratio.}
 		\label{tab:contpar}
 	\end{minipage}
 	\end{turn} 	
\end{table*} 
  
\section{Summary}
\label{sec:summary}

We have presented a comprehensive atlas of composite spectra of SDSS galaxies. \textcolor{black}{The signal-to-noise ratio of the composites can reach as high as $ S/N \simeq 132 - 4760 $ at {$ \Delta \lambda = 1 $ \AA} sampling. We have shown that the volume limited sampling is very important to get physically meaningful composites at very short and very long wavelength. We have argued that our robust averaging method is superior to both traditional averaging and median calculation in terms of computational efficiency and physical interpretation.} We have calculated the most important spectral feature indicators of the composites and shown that the parameters of the spectra span almost as wide a range as the spectral parameters of individual galaxies do. This makes the atlas a representative collection of the numerous types of galaxies observed by SDSS.

The composite spectrum atlas is available online at \url{http://www.vo.elte.hu/compositeatlas}.

\section*{Acknowledgments}

This work was supported by the following Hungarian grants: NKTH: Pol\'anyi, OTKA-80177 and KCKHA005.

The Project is supported by the European Union and co-financed by the European Social Fund (grant agreement no. T\'AMOP 4.2.1./B-09/1/KMR-2010-0003).

\label{lastpage}

\end{document}